\documentclass[preprint,review,12pt]{elsarticle}

\usepackage[colorlinks,
            linkcolor=blue,
            anchorcolor=blue,
            citecolor=blue]{hyperref}

\usepackage{amssymb,graphicx,float,multirow,longtable,array,amsmath,caption,subcaption,arydshln}

\usepackage{lineno}
\nolinenumbers
\usepackage{graphicx}
\usepackage{multirow}
\usepackage{tabularx}
\usepackage{xcolor}
\usepackage{booktabs}
\usepackage{upgreek}
\usepackage{pifont}
\usepackage{enumitem}

\begin{document}

\begin{frontmatter}



\title{A Closed-Form Design Method for U-Shaped Springs in Aeroelastic Modeling of Truss-Girder Suspension Bridges}


\author[inst1]{Guangzhong Gao\corref{cor1}}
\ead{ggz@chd.edu.cn}
\author[inst1]{Wenkai Du}
\ead{dwk@chd.edu.cn}
\author[inst2]{Yanbo Sun}
\ead{yanbosun1999@163.com}
\author[inst1]{Yonghui Xie}
\ead{854060570@chd.edu.cn}
\author[inst1]{Jiawu Li}
\ead{ljw@gl.chd.edu.cn}
\author[inst3,inst4,inst5]{Ledong Zhu}
\ead{ledong@tongji.edu.cn}

\affiliation[inst1]{organization={Highway College},
            addressline={Chang'an University}, 
            city={Xi'an},
            postcode={710064}, 
            state={Shaanxi},
            country={China}}

\affiliation[inst2]{organization={Xi'an Rail Transit Group Company Limited},
            city={Xi'an},
            postcode={330200}, 
            state={Shaanxi},
            country={China}}

\affiliation[inst3]{organization={Department of Bridge Engineering},
            addressline={Tongji University}, 
            postcode={200092}, 
            state={Shanghai},
            country={China}}
\affiliation[inst4]{organization={State Key Laboratory of Disaster Reduction in Civil Engineering},
            addressline={Tongji University}, 
            postcode={200092}, 
            state={Shanghai},
            country={China}}
\affiliation[inst5]{organization={Key Laboratory of Transport Industry of Bridge Wind Resistance Technology},
            addressline={Tongji University}, 
            postcode={200092}, 
            state={Shanghai},
            country={China}}

\cortext[cor1]{Corresponding author: Guangzhong Gao, Associate Professor, Chang'an University, 710064, Xi'an, Shaanxi, China.}

\begin{abstract}

Aeroelastic model testing is essential for evaluating the wind resistance performance of long-span suspension bridges. In these models, truss girders are commonly modelled by a discrete-stiffness system incorporating U-shaped springs to simulate elastic stiffness properties, including vertical bending, lateral bending, and torsional rigidity. The design of these U-shaped springs significantly influences the accuracy of aeroelastic model tests for truss-girder suspension bridges. Traditional design approaches, however, lack an analytical foundation, relying instead on trial-and-error searches across the full parameter space, which is computationally intensive and time-consuming. 

To overcome these limitations, this study introduces a novel design method that establishes closed-form equations for U-shaped spring design. The method simplifies the truss girder's supporting condition to a cantilever configuration and determines the corresponding elastic stiffness using the principle of elastic-strain energy equivalence. These closed-form equations transform the design process into a non-smooth optimization problem, accounting for precision constraints inherent in practical fabrication. Derivative-free optimization algorithms, including the Nelder-Mead method, Pattern Search method, and Genetic Algorithm, are employed to identify a global optimal solution.

The proposed method is validated through its application to a representative suspension bridge with a truss girder, with numerical and experimental results confirming its accuracy and reliability. This approach enhances aeroelastic model design techniques for long-span bridges by providing a theoretical framework for the design of U-shaped spring, reducing the optimization process to several seconds. Moreover, the proposed design method can be extended to the aeroelastic modeling of other truss-based structures, such as transimission towers and arch bridges.
\end{abstract}


\begin{highlights}
\item Closed-form design method of U-shaped springs for truss-girder aeroelastic models. 
\item Non-smooth optimization reduces design time from hours to seconds.
\item Method validated numerically and experimentally for truss-girder bridge aeroelasticity.
\item Potential application to other truss-type structures such as transmission towers and arch bridges. 
\end{highlights}

\begin{keyword}
Aeroelastic modeling \sep Truss-girder suspension bridges \sep U-shaped spring \sep Closed-form design \sep Non-smooth optimization
\end{keyword}

\end{frontmatter}


\section{Introduction}
\label{sec:Introduction}

Truss girders are widely utilized in long-span bridges as a fundamental structural component. They are particularly well-suited to mountainous regions, where transporting large prefabricated beam segments, such as concrete or steel box girders, presents significant challenges. Truss girders can be fabricated into smaller modular units, facilitating easier transportation and on-site assembly. However, their bluff aerodynamic configuration renders them vulnerable to various wind-induced vibrations, including vortex-induced vibration (VIV) and flutter. Numerous long-span truss-girder bridges, such as the Golden Gate Bridge \citep{GoldenGate2025}, Akashi Kaikyo Bridge \citep{miyata2017akashi,miyata1993aerodynamics}, Verrazzano-Narrows Bridge \citep{Leaden2020}, and Yangsigang Yangtze River Bridge \citep{sun2016experimental}, have experienced these vibrations, necessitating the adoption of structural or aerodynamic countermeasures. Consequently, in the design of long-span truss-girder bridges, accurately evaluating and optimizing their aerodynamic performance is critical to ensuring structural safety and stability \citep{yousaf2021flutter}.

Wind tunnel experiments are fundamental tools in investigating complex wind-structure interactions for bluff bridge components. The use of wind tunnel tests for flexible bridges dates back to the 1940s, pioneered by F. Burt Farquharson during his investigation of the original Tacoma Narrows Bridge \citep{university1949aerodynamic}. Farquharson employed two primary types of wind tunnel tests---section model tests and full bridge aeroelastic model tests---which have since become the most widely adopted experimental techniques in bridge wind engineering \citep{maheux2022theory}.

Section model tests are relatively simple and can be conducted at larger scales, thereby minimizing the Reynolds number effect and preserving geometric details. These tests, typically involving elastically supported section models, are conventionally used for aerodynamic shape optimization, identification of aerodynamic parameters, and observation of nonlinear aerodynamic phenomena \citep{maheux2022theory,gao2022nonlinear}. However, section model tests have limitations, including their inability to capture structural nonlinearities \citep{maheux2022theory}, multi-mode coupling effects \citep{gao2025design}, the nonlinear influence of additional wind angles of attack \citep{zhang2025effects}, and the complex wind fields associated with mountainous terrain \citep{yousaf2021flutter,li2024advances,shen2021nonuniform}. Consequently, section model tests are often paired with full aeroelastic model tests \citep{diana2013wind,liu2017effect}. The former is typically applied during the conceptual design stage for preliminary assessments, while the latter is utilized in the final design stage for comprehensive evaluations of various aerodynamic performances.

\begin{figure}[H]
    \centering
\includegraphics[width=1\textwidth]{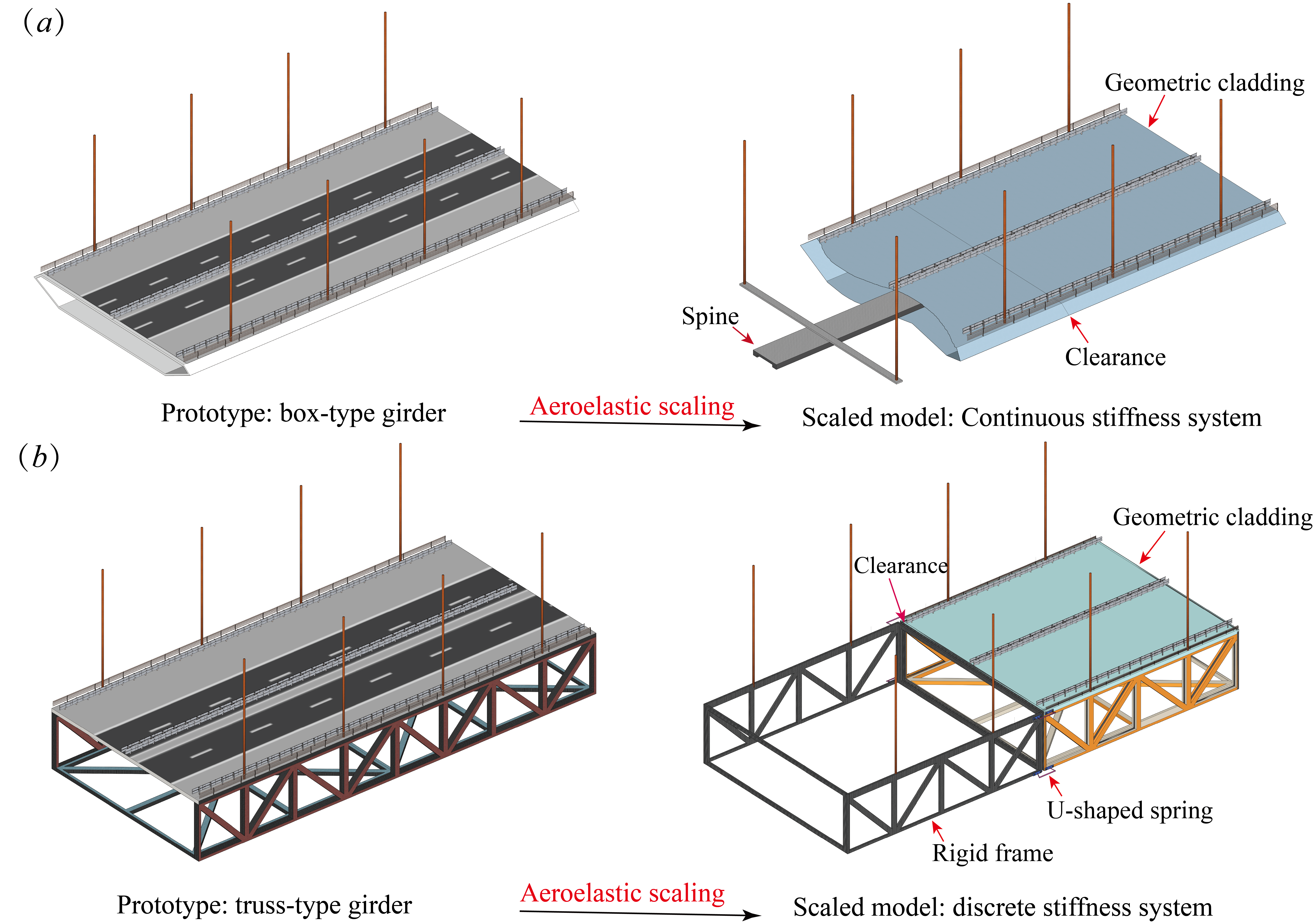}
    \caption{Schematic illustration of aeroelastic modeling techniques for typical bridge decks. (\textit{a}) Continuous stiffness system for box-type girders, (\textit{b}) Discrete stiffness system for truss-type girders.}
    \label{fig:1}
\end{figure}

The design of a full bridge aeroelastic model is crucial for ensuring the accuracy of wind tunnel test results. Full bridge aeroelastic models must satisfy scaling requirements for elastic stiffness, mass properties, and geometric shape, in accordance with Froude number similarity for cable-supported bridges \citep{irwin2017full}. Fig. \ref{fig:1} illustrates two common techniques for the aeroelastic modeling of bridge girders: (\textit{a}) the continuous stiffness system for box-type girders, and (\textit{b}) the discrete stiffness system for truss-type girders. 

In the continuous stiffness system (see Fig. \ref{fig:1}\textit{a}), the elastic stiffness of the girder is simulated by a metal spine beam that extends continuously along the entire length of the model. This beam is positioned at the shear center and connected to hangers or cables via horizontal metal beams. Geometric claddings are mounted onto these beams to ensure that the aerodynamic configuration aligns with that of the prototype bridge. These claddings, made of lightweight but stiff materials, are designed solely to maintain geometric similarity and do not contribute to structural stiffness. Consequently, they are typically divided into discrete segments with clearances to accommodate the flexible deformation of the metal spine. Additional mass blocks are placed within the geometric claddings to meet the scaling requirements for the mass and mass moment of inertia of the girder. This continuous stiffness system is widely utilized for the aeroelastic modeling of box-type girders \citep{zhu2007flutter,diana1995comparisons,li2024buffeting,ma2022design,ge2018full}, as the spine beam can be positioned inside the cladding at the shear center without altering the external aerodynamic shape.

The continuous stiffness system is not suitable for truss-type girders because all truss members are exposed to wind flow, and an internal spine beam positioned at the shear center would alter the aerodynamic shape. One compromise is to place a continuous spine beam beneath the bridge deck, within the wake of the upper chord members; however, this leads to misalignment of the stiffness center. A more feasible solution is to adopt a discrete-stiffness system utilizing U-shaped springs. As depicted in Fig. \ref{fig:1}\textit{b}, this system employs four U-shaped springs to simulate the flexural stiffness of the truss girder, with geometric claddings replicating the scaled shape. These claddings are reinforced by interior rigid frames to provide reliable support for hangers or cables and are connected end-to-end across clearances via U-shaped springs. The springs are mounted on the upper and lower chord members and positioned toward the shear center. Consequently, the simulation of elastic stiffness in aeroelastic models of truss girders is reduced to the design of these U-shaped springs \citep{xu2009long,yu2017research,song2025investigation}.

Another significant challenge in designing aeroelastic models for truss girders is satisfying the scaled mass properties. For instance, in the Huajiang Canyon Bridge—a typical truss-girder suspension bridge (see Section \ref{sub:a_long_span_suspension_bridge_with_a_truss_girder})—the design value for a 17.8 cm long girder segment is only 215 g. This small mass can be easily exceeded when ensuring the stiffness of the rigid frames and geometric cladding. In a continuous stiffness system, the mass of the metal spine---often designed with solid sections such as channels or T-sections for manufacturing convenience---constitutes a substantial portion of the total mass and cannot be readily reduced. In contrast, the discrete-stiffness system using U-shaped springs offers an advantage in mass simulation due to the minimal mass of the springs. Additional mass blocks can be placed on the lower side of the bridge deck to achieve accurate mass scaling.

The concept of employing a discrete-stiffness system in aeroelastic modeling of truss girders was pioneered by Miyata et al. \citep{miyata2017akashi,miyata1993aerodynamics} during the aeroelastic model testing of the Akashi Kaikyo Bridge. Their model utilized V-shaped springs to connect truss-girder segments. Their sectional model tests, conducted with and without these springs, demonstrated that V-shaped springs had only a minor impact on the aerodynamic properties of the truss girder. Over time, V-shaped springs evolved into U-shaped springs to incorporate additional parameters for simulating sectional stiffness \citep{xu2009long}. More recently, Lan et al. \citep{lan2024new} introduced a multi-spine frame system for double-deck steel truss girders. This system employs three longitudinal spine beams, connected by thin horizontal beams and vertical columns, to form a rigid frame. The longitudinal spine beams are concealed beneath the bridge decks to minimize aerodynamic interference while satisfying mass property scaling and stiffness center alignment. However, this approach is limited to double-deck truss girders, whereas the discrete-stiffness system with U-shaped springs remains versatile across a broader range of truss configurations.

Similar challenges in simulating flexural stiffness arise in the aeroelastic modeling of other large-scale truss structures, such as transmission towers and arch bridges \citep{liang2017an,li2021aeroelastic}. U-shaped springs have also been applied to the aeroelastic modeling of truss-type transmission towers by Liang et al. \citep{liang2017an}.

Despite over three decades of practical application, the design method for U-shaped springs in aeroelastic modeling remains underdeveloped. The scarcity of literature on this topic may stem from the complexity introduced by the multiple geometric parameters of U-shaped springs and the complex structural behavior of truss-girder suspension bridges, which complicates the development of a closed-form design method. As reviewed by Yu \citep{yu2017research}, conventional design approaches for U-shaped springs rely on a trial-and-error procedure, typically requiring approximately $10^7$ batches of finite element calculations to explore the parameter space and identify optimal design parameters. This direct parameter search is computationally intensive and time-consuming.

This study addresses these limitations by proposing a novel design procedure that simplifies supporting conditions and derives closed-form design equations based on the principle of equivalent elastic-strain energy. Optimal geometric parameters are determined using well-established global optimization methods, with consideration given to geometric constraints imposed by manufacturing precision. The optimal design parameters for U-shaped springs, which satisfy the scaling of elastic stiffness for truss girders, can be obtained in seconds. Theoretical results are validated using the finite element method and wind tunnel experiments of a typical suspension bridge. The proposed design method is extendable to the aeroelastic modeling of truss-type arch bridges and transmission towers.

The remainder of the paper is structured as follows. Section \ref{sec:design_method_for_aeroelastic_models_of_truss_type_girders} outlines the general scaling principles of aeroelastic modeling, the detailed setup and design procedure of the discrete-stiffness system, and the identification of elastic stiffnesses of truss girders through elastic strain energy equivalence. Section \ref{sec:derivation_of_design_equations_for_the_u_shaped_springs} presents the derivation of closed-form design equations for U-shaped springs. Section \ref{sec:optimization_of_design_parameters_and_numerical_validation} proposes an optimization method for determining the optimal geometric parameters of U-shaped springs, with numerical validation via finite element analysis in terms of modal frequencies and aeroelastic deformation. Section \ref{sec:experimental_case_study} provides an experimental case study, detailing the fabrication of an aeroelastic model of a typical suspension bridge using the proposed design method and quantifying sources of error in fabrication and installation. Section \ref{sec:conclusions} summarizes the conclusions and discussions.

\section{Design Method for aeroelastic models of truss-type girders} 
\label{sec:design_method_for_aeroelastic_models_of_truss_type_girders}

\subsection{Scaling principles} 
\label{sub:scaling_principles}

In the design of full-bridge aeroelastic models, ensuring geometric similarity is critical, as aerodynamic forces acting on bridge components are directly governed by their geometric configurations. To accurately replicate wind-induced responses, the modal parameters of the major lower-order vibration modes must be accurately reproduced, necessitating similarity in both elastic stiffness and mass properties. For a detailed examination of scaling requirements, readers are referred to \citep{irwin2017full}.

For suspension bridges, gravity significantly influences the structural stiffness and thus Froude number similarity should be satisfied. The Froude number is defined as ${\rm{\text{Fr}}} = U/\sqrt {gL}$, where $U$ represents the local wind velocity, $g$ is the gravitational acceleration, and $L$ denotes the characteristic length. The Froude scaling dictates the velocity scaling $\lambda_U$ to be the square root of the length scaling  $\lambda_L$, i.e., $\lambda_U = \lambda_L^{0.5}$.

The simulation of mass properties requires that the mass, mass moment of inertia, and center of gravity conform to the following scaling ratios:
\begin{equation}
    \label{eq:1}
{m_{\rm{m}}} = \lambda _L^2{m_{\rm{p}}},{I_{\rm{m}}} = \lambda _L^5{I_{\rm{p}}},{\left( {\frac{{{y_{c,g}}}}{D}} \right)_{\rm{m}}} = {\left( {\frac{{{y_{c,g}}}}{D}} \right)_{\rm{p}}}
\end{equation}
where $m_{\rm{m}}$ and $m_{\rm{p}}$ represent the mass per unit length of the aeroelastic model and prototype bridge, repesectively. $I_{\rm{m}}$ and $I_{\rm{p}}$ denote the mass moment of inertia per unit length of the model and prototype bridge, repesectively. $y_{c,g}$ indicates the vertical position of gravity center, and $D$ is the height of the truss girder.

The elastic stiffness includes the vertical bending rigidity $EI_z$ , the transverse bending rigidity $EI_y$ and Saint-Venant free torsional rigidity $GJ_\text{d}$. In long-span truss girders, the lower-chord members constrain out-of-plane warping, as a result, the Saint-Venant torsional stiffness $GJ_\text{d}$ is the primary contributor to torsional response, while the warping (constrained) torsional stiffness $EI_{\omega}$ is negligible. 

Adhering to Froude number similarity, the scaling ratios of these rigidity parameters are expressed as
\begin{equation}
    \label{eq:2}
{\left( {E{I_z}} \right)_{\rm{m}}} = \lambda _L^5{\left( {E{I_z}} \right)_{\rm{p}}},{\left( {E{I_y}} \right)_{\rm{m}}} = \lambda _L^5{\left( {E{I_y}} \right)_{\rm{p}}},{\left( {G{J_\text{d}}} \right)_{\rm{m}}} = \lambda _L^5{\left( {G{J_\text{d}}} \right)_{\rm{p}}}
\end{equation}
where subscript $\rm{m}$ refers to the aeroelastic model, and $\rm{p}$ denotes the prototype bridge.The length scaling $\lambda_L$ typically falls within the range of 1:100 to 1:300.


\subsection{Discrete stiffness system and design procedure of U-shaped springs} 
\label{sub:discrete_stiffness_system_and_design_procedure_of_u_shaped_springs}

\begin{figure}[H]
    \centering
\includegraphics[width=1\textwidth]{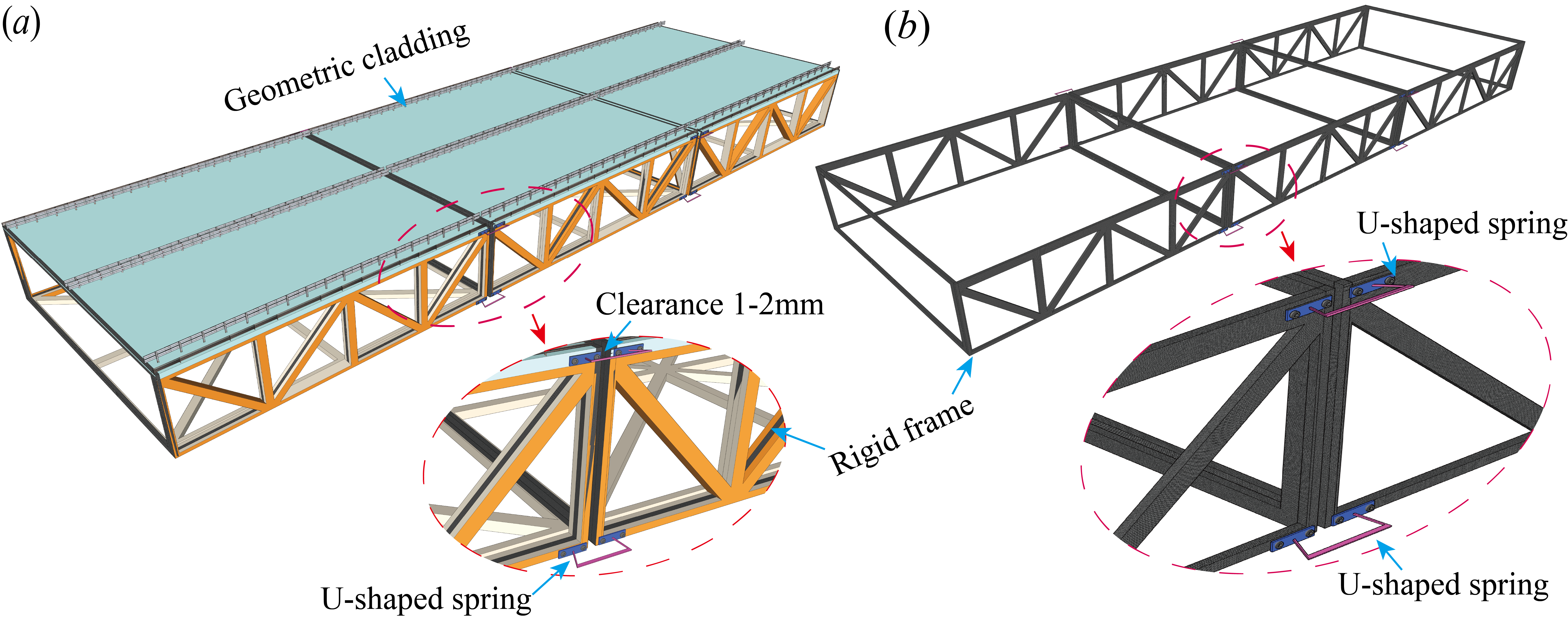}
    \caption{Conceptual description of the aeroelastic model of truss-type girder using a discrete stiffness system. (\textit{a}) General setup of assembled three segments, (\textit{b}) Three rigid frames are connected end-to-end across the clearance by U-shaped springs.}
    \label{fig:2}
\end{figure}

The aeroelastic model of truss girders, using a discrete stiffness system, is composed of rigid frames, geometric claddings, and U-shaped springs, as illustrated in Fig. \ref{fig:2}. Each scaled girder segment consists of an internal rigid frame and external geometric cladding. The geometric cladding is fabricated from lightweight but stiff materials, such as wood or foam, and is attached to the rigid frames to provide the scaled aerodynamic shape of the truss girder. The rigid frames are constructed from carbon fiber or metal plates, integrally engraved and rigidly assembled to ensure overall rigidity. These frames provide anchor points for the attachment of U-shaped springs and hangers. The rigid frames must possess sufficient stiffness to prevent any distortion that could result in misalignment or friction with the attached cladding. Adjacent rigid frames are connected end-to-end by four U-shaped springs, positioned at locations corresponding to the truss chords, spanning a clearance of 1 to 2 mm. These clearances are typically placed at the midpoints of hanger intervals, with each segment occupying an effective length of 1-2 hanger intervals.

The four U-shaped springs are directed toward the shear center of the truss section. Vertical and lateral bending of the truss girder are simulated through the axial extension and contraction of U-shaped springs at the clearances, while torsion is simulated via out-of-plane relative movement of these springs, enabling twisting between adjacent segments. 

\begin{figure}[H]
    \centering
\includegraphics[width=1\textwidth]{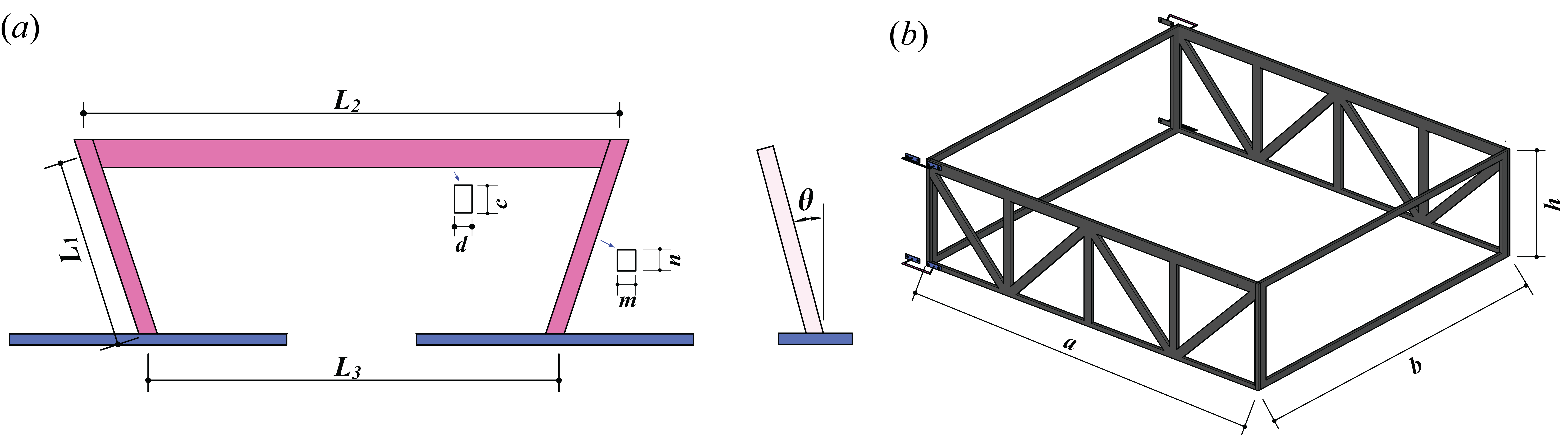}
    \caption{Geometric variables of the discrete stiffness system. (\textit{a}) U-shaped spring, (\textit{b}) Rigid-frame block connected to four U-shaped springs at the corners.}
    \label{fig:3}
\end{figure}

For simplicity, the same set of U-shaped springs is used to simulate the stiffness properties of the truss girder. Four U-shaped springs are arranged symmetrically around the vertical centerline of the girder, positioned at the four truss chords. Each U-shaped spring is oriented with its plane directed toward the shear center of the truss girder. As shown in Fig. \ref{fig:3}\textit{a}, the U-shaped springs are designed to exhibit symmetry about their local vertical axis, leading to the following geometric parameters:
\begin{equation}
    \label{eq:3}
\boldsymbol{\rm{\upmu}} = {[\begin{array}{*{20}{c}}
m&n&c&d&{{L_1}}&{{L_2}}&{{L_3}}
\end{array}]^{\rm{T}}}
\end{equation}
where $m$ and $n$ represent the in-plane and out-of-plane dimensions of the column cross-section of the U-shaped spring, respectively; $c$ and  $d$ denote the in-plane and out-of-plane dimensions of the crossbeam cross-section; $L_1$ and  $L_2$ indicate the height of the column and the length of the crossbeam, respectively. $L_3$ signifies the distance between the fixed ends of the two columns.

As shown in Fig. \ref{fig:3}\textit{b}, the geometric parameters of a rigid frame include the axial length $a$, width $b$ and height $h$. These parameters are determined based on the length scale and dimensions of the prototype bridge. Specifically, $a$ is the scaled length of discrete segment minus the clearances at both ends. A standard discrete segment typically spans 1 to 2 hanger intervals. The width $b$ and height $h$ are determined by the girder width and height after subtracting the geometric cladding, as indicated by Fig. \ref{fig:1}\textit{b}.

The geometric parameters of the U-shaped springs are selected to ensure that the stiffness properties of the truss girder are precisely simulated by the discrete stiffness system. Consequently, the design equations for the U-shaped springs are formulated as functions of the geometric parameter vector:
\begin{equation}
    \label{eq:4}
\left( E I_z \right)_{\mathrm{m}} = f_1 \left( \boldsymbol{\upmu} \right), \left( E I_y \right)_{\mathrm{m}} = f_2 \left( \boldsymbol{\upmu} \right), \left( G J_\text{d} \right)_{\mathrm{m}} = f_3 \left( \boldsymbol{\upmu} \right)
\end{equation}

These parameters are subject to constraints, including lower and upper bounds, and a manufacturing precision limitation that restricts variations to discrete increments, such as integer multiples of 0.1 mm. These constraints are expressed as:
\begin{equation}
    \label{eq:5}
{\boldsymbol{\upmu }} = 0.1 \times {\bf{k}},\;{\bf{k}} \in {\mathbb{Z}^6},\;{k_{i,\min }} \le {k_i} \le {k_{i,\max }},\;i = 1,2, \ldots ,6
\end{equation}
where ${k_{i,\min }}$ and ${k_{i,\max }}$ represent the lower and upper bounds of each geometric parameter, respectively. 

Therefore, the design of U-shaped springs reduces to a constraint optimization problem described by Eqs. (\ref{eq:4})-(\ref{eq:5}). The geometric parameters are determined by minimizing the residuals in the design equations. The challenge is to determine the functional form of Eq. (\ref{eq:4}). The proposed design procedure is outlined as follows:

\textbf{Step I}: Simplify the truss girder's supporting condition from an elastically supported configuration at multiple hanger points to a cantilever scenario.

\textbf{Step II}: Determine the equivalent stiffnesses $EI_z$, $EI_y$ and $GJ_d$ for the prototype bridge using the principle of elastic-strain-energy equivalence for the cantilever truss girder, as detailed in Section \ref{sub:equivalent_euler_bernoulli_beam_by_elastic_strain_energy_equivalence}.

\textbf{Step III}: Derive the stiffness ${\left[ {{{\bf{K}}_{\rm{u}}}} \right]^{\rm{e}}}$ of a cantilever U-shaped spring element in Section \ref{sub:stiffness_matrix_of_u_shaped_spring_element}, and transform it into the stiffness matrix ${\left[ {\bf{K}} \right]^{\rm{e}}}$ of a cantilever super-element comprising four U-shaped springs and a rigid frame, which is in Section \ref{sub:stiffness_matrix_of_a_cantilever_spring_lever_element}.

\textbf{Step IV}: Establish the functional form of the design equations via the equivalence of elastic-strain-energy between the cantilever Euler-Bernoulli beam and discrete stiffness model, which is addressed in Section \ref{sub:design_equations_by_elastic_strain_energy_equivalence}. 

\textbf{Step V}: Optimize the design equations to determine the optimal geometric parameters of the U-shaped springs, as presented in Section \ref{sub:optimization_algorithm}.

\textbf{Step VI}: Validate the accuracy of the design parameters through numerical and experimental tests on the full-bridge aeroelastic model, as reported in Sections \ref{sub:numerical_validation_of_modal_frequencies}-\ref{sub:comparison_of_aerostatic_deformation} and Section \ref{sec:experimental_case_study}.

\begin{figure}[H]
    \centering
\includegraphics[width=1\textwidth]{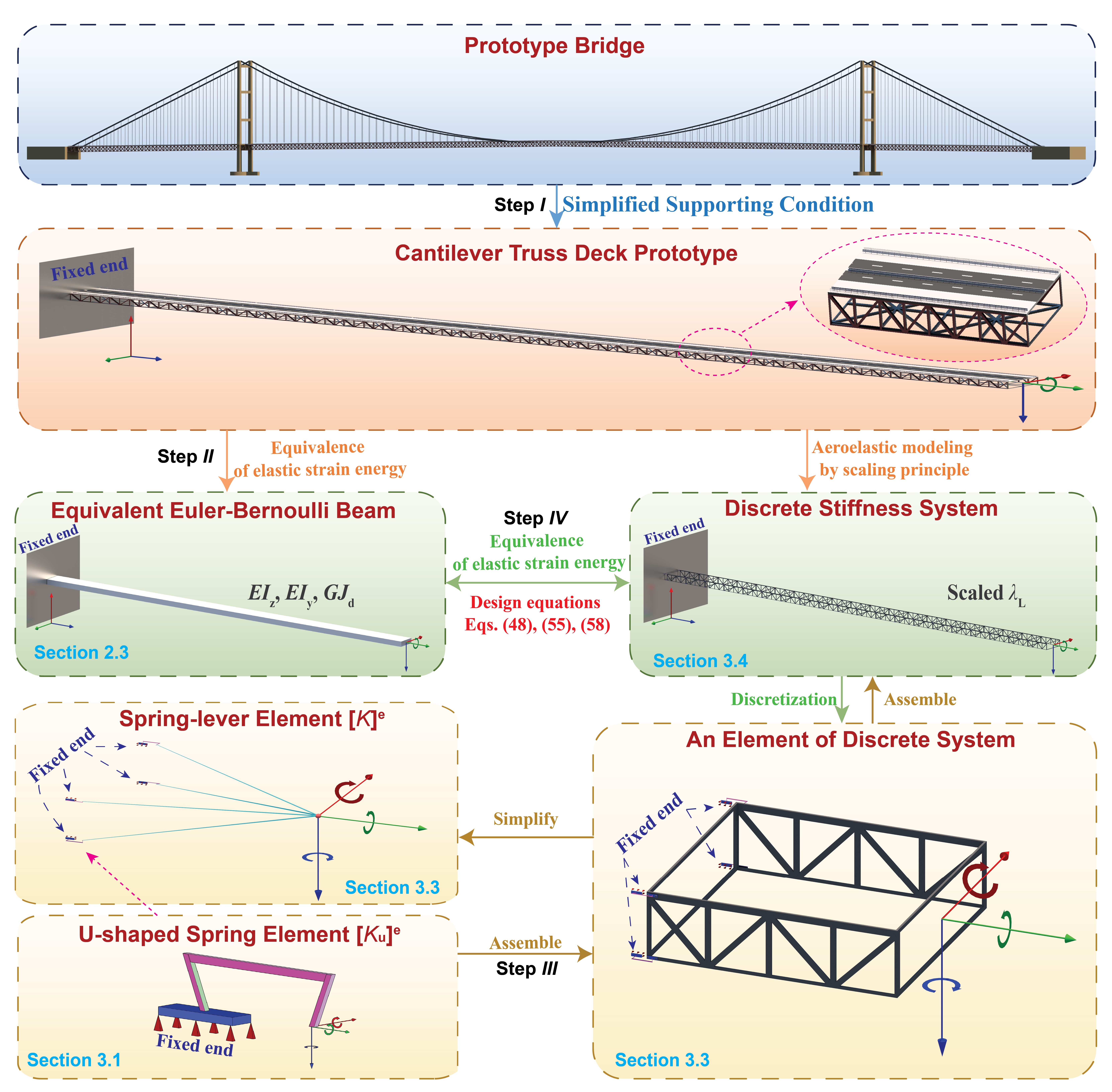}
    \caption{General design procedure for the discrete stiffness system with U-shaped springs.}
    \label{fig:4}
\end{figure}

\subsection{Equivalent Euler-Bernoulli beam by elastic-strain-energy equivalence} 
\label{sub:equivalent_euler_bernoulli_beam_by_elastic_strain_energy_equivalence}

As specified by Eq. (\ref{eq:2}), the bending moments of inertia $I_z$, $I_y$ along with the torsional constant $J_\text{d}$ of the prototype truss girder are critical parameters for designing the aeroelastic model. The principle of elastic-strain-energy equivalence is employed to obtain an equivalent Euler-Bernoulli beam representation of the prototype truss girder. This approach involves equating the elastic energy stored during deformation when an external load is applied to the free end of a cantilever truss girder. Assuming a quasi-static deformation process devoid of vibration, the elastic-strain energy equals the work performed by the external force. Therefore, the equivalence of elastic-strain-energy simply requires that the cantilever prototype truss girder and its equivalent Euler-Bernoulli beam exhibit identical deflections at the free end under the same span length and external load conditions. To facilitate this, a finite element model of the cantilever truss girder is initially developed.

The vertical bending moment of inertia $I_z$ is determined by analyzing the vertical deflection resulting from a vertical load applied to the free end of the cantilever prototype truss girder along the cross-section's centerline, as computed using the finite element model. For an Euler-Bernoulli beam, the vertical deflection at the free end is expressed as:
\begin{equation}
    \label{eq:6}
{v_y} = \frac{{Fl_x^3}}{{3E{I_z}}}
\end{equation}
where ${v_y}$ is obtained directly from the finite element model of the prototype truss girder. $F$ is the applied vertical load. $E$  is the Young's modulus. $l_x$ is the axial length of the truss girder. $I_z$ is the bending moment of inertia around the horizontal axis, to be determined for the corresponding Euler-Bernoulli beam to satisfy the equivalence of elastic-strain-energy, and it can be obtained according to Eq. (\ref{eq:6}) as
\begin{equation}
    \label{eq:7}
{I_z} = \frac{{Fl_x^3}}{{3E{v_y}}}
\end{equation}

The torsional costant $J_\text{d}$ is deduced from the torsional angle $\theta$  at the free end when a torque $T$ is applied, incorporating the effects of warping rigidity \citep{yue2022research}. The equivalent torsional constant can be calculated as
\begin{equation}
    \label{eq:8}
J_\text{d}^ *  = \frac{{Tl_x^{}}}{{G\theta }}
\end{equation}
where $J_\text{d}^ *$ is the equivalent torsional constant that incorporates the effect of warping constraints.

\begin{figure}[H]
    \centering
\includegraphics[width=0.6\textwidth]{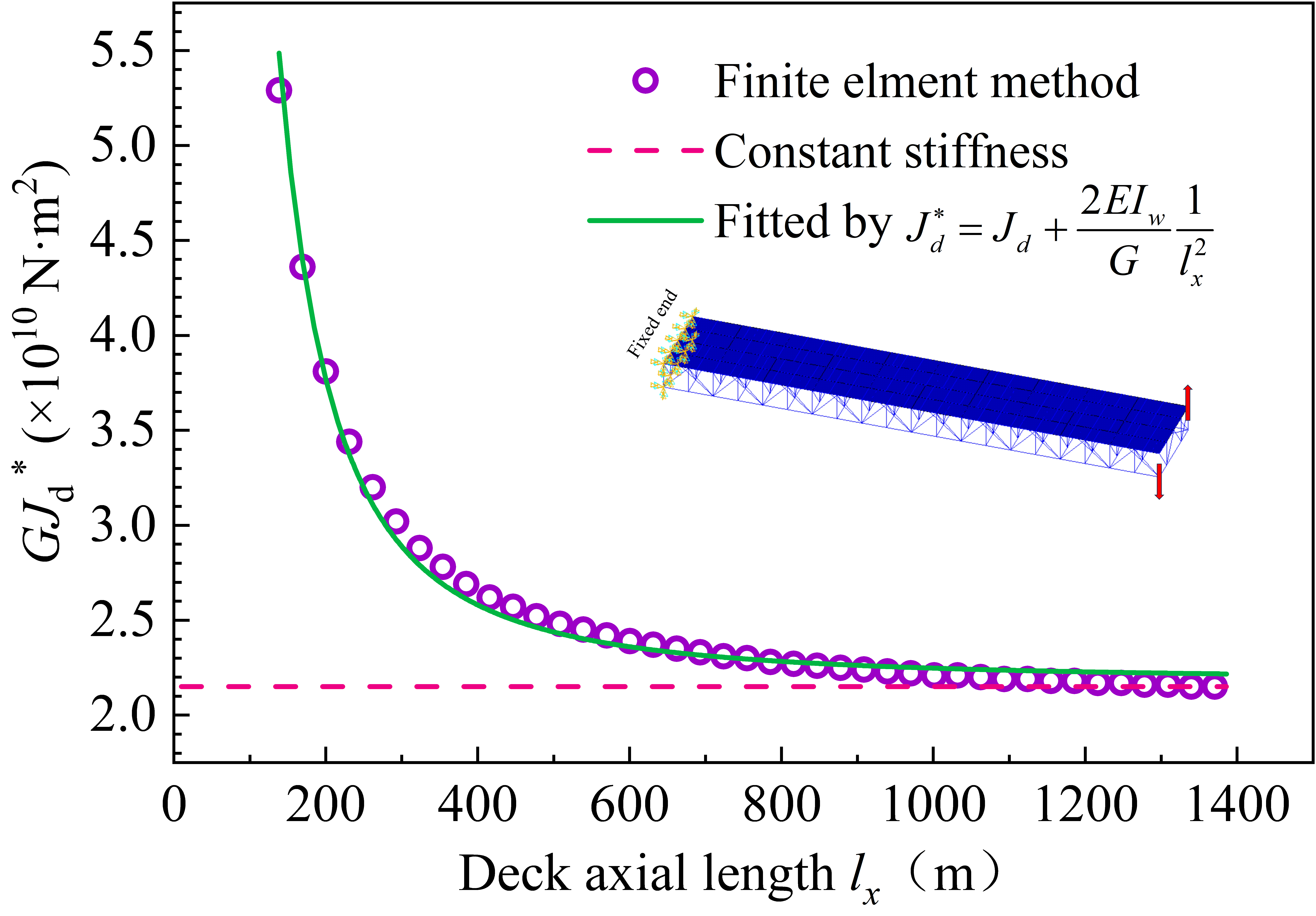}
    \caption{Identification of torsional constant of the truss girder. }
    \label{fig:5}
\end{figure}

However, warping constraints $J_\text{d}^ *$ cause to vary with the girder's length.  As shown in Fig.~\ref{fig:5}, the torque is applied via a couple of vertical forces on the vertical chords. Significant variation in the resulting $J_\text{d}^ *$ is observed due to the warping rigidity of the main truss. The equivalent torsional constant can be obtained by expressing the torsional angle $\theta$ using the vertical deflection as $\theta = 2v_y/B_*$ and equivating the torque at any section using $E{I_{\omega} }\theta ''' + G{J_\text{d}}\theta ' = GJ_\text{d}^ * \theta '$ , noting that $v_y$ is a function of the length $l_x$ as in Eq. (\ref{eq:6}). The equivalent torsional constant, accounting for warping, is given by
\begin{equation}
    \label{eq:9}
J_\text{d}^ * \left( {{l_x}} \right) = {J_\text{d}} + \frac{{2E{I_\omega }}}{G}\frac{1}{{l_x^2}}
\end{equation}
where $J_\text{d}$ is the torsional constant of the truss girder. $G$ is the shear modulus. $I_{\omega}$ represents the warping constant.

Fig.~\ref{fig:5} illustrates the fitting of torsional constant using Eq. (\ref{eq:9}) against finite element method calculations. It is found that Eq. (\ref{eq:9}) can capture the variation of $J_\text{d}^ *$ with axial length $l_x$. The torsional constant of the truss girder $J_\text{d}$ can be obtained from the fitted formula of Eq. (\ref{eq:9}).

\begin{figure}[H]
    \centering
\includegraphics[width=0.95\textwidth]{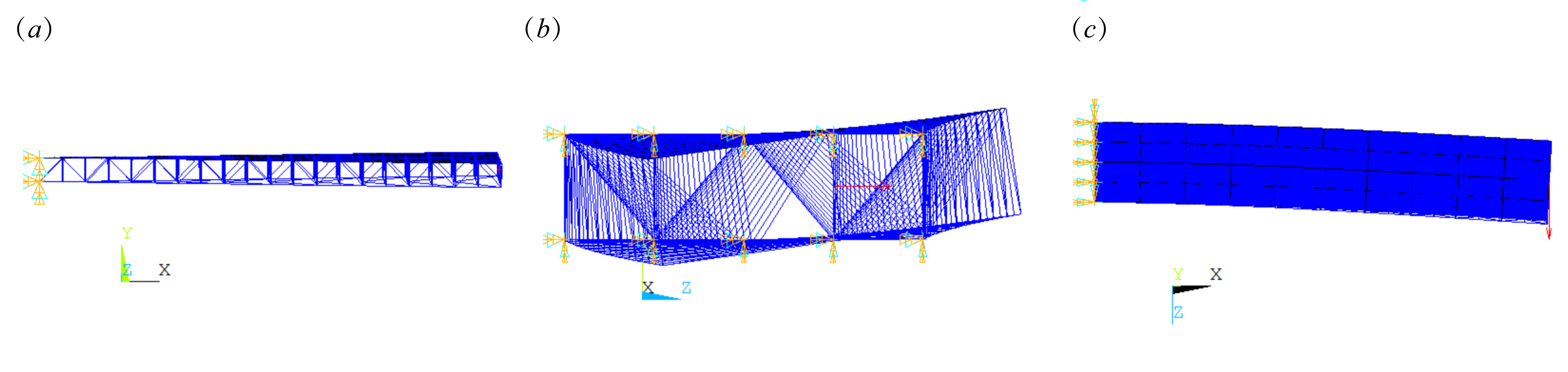}
    \caption{Correction of lateral-torsion coupling for the equivalent transverse stiffness. (\textit{a}) Elevation view, (\textit{b}) Cross-section view, (\textit{c}) Top view.}
    \label{fig:6}
\end{figure}

The transverse bending moment of inertia $I_y$  is calculated by applying a transverse load to the free end of the cantilever truss girder through its shear center to prevent coupled torsional deformation. Since the shear center's location is not known a priori, its vertical position along the line of vertical symmetry is first established. Initially, a transverse force  $F_z$ is applied at the geometric center, inducing coupled torsion-bending deformation (see Fig.~\ref{fig:6}). The offset distance to the shear center is then derived from the torsional deformation $\theta$  as  ${e_y} = GJ_\text{d}^ * \theta /\left( {{F_z}{l_x}} \right)$. Subsequently, by adjusting the application point of  $F_z$ by $e_y$, pure transverse bending deformation is achieved. The transverse bending moment of inertia $I_y$ is then computed using:
\begin{equation}
    \label{eq:10}
{I_y} = \frac{{{F_z}l_x^3}}{{3E{v_z}}}
\end{equation}

The equivalent mass moment of inertia of the truss girder is determined using a simplified method proposed by Hua et al. \citep{hua2017a}. This method involves attaching additional mass moment of inertia on the truss nodes of a cantilevere girder in the finite element model, and calculating the mass moment of inertia based on the change in the torsional frequency, expressed as
\begin{equation}
    \label{eq:11}
{I_m} = \frac{{\Delta {I_m}{{\left( {f_{{\rm{t,}}1}^*} \right)}^2}}}{{l\left[ {{{\left( {{f_{{\rm{t,}}1}}} \right)}^2} - {{\left( {f_{{\rm{t,}}1}^*} \right)}^2}} \right]}}
\end{equation}
where $I_m$ is the equivalent mass moment of inertia per unit length. $l$ is the axial length of each truss girder segment. $\Delta {I_m}$ represents the attached mass moment of inertia on each truss segment. ${f_{{\rm{t,}}1}}$, ${f_{{\rm{t,}}1}^*}$ denote the first-order torsional frequency of the cantilever steel truss girder before and after adding the mass moment, respectively.


\section{Derivation of design equations for the U-shaped springs} 
\label{sec:derivation_of_design_equations_for_the_u_shaped_springs}

In this section, we present the derivation of the design equations for U-shaped springs within a discrete stiffness system. To streamline this process, the elastically-supported truss girder of the prototype bridge is approximated as a cantilever structure. This simplification enables a more tractable analytical framework for deriving the stiffness properties of the U-shaped springs. As depicted in Fig.~\ref{fig:4}, the cantilever truss girder is discretized into a series of rigid-frame elements. Each element comprises a rigid frame with four U-shaped springs attached at designated points, typically along the truss chords. For analytical simplicity, each rigid-frame element is further modeled as a ``spring-lever element'', which integrates four U-shaped springs with rigid beams. This spring-lever element is subsequently discretized as individual cantilever U-shaped spring elements connected to rigid beams.

The stiffness matrix derivation is structured as follows: Section \ref{sub:stiffness_matrix_of_u_shaped_spring_element} addresses the derivation of the stiffness matrix for a single U-shaped spring element in its local coordinate system. Section \ref{sub:cantilever_u_shaped_spring_element_with_a_rigid_beam} extends to the stiffness matrix of a cantilever U-shaped spring with a rigid beam in the global coordinate system. Section \ref{sub:stiffness_matrix_of_a_cantilever_spring_lever_element} presents the assembly of the stiffness matrix for the spring-lever element. Section \ref{sub:design_equations_by_elastic_strain_energy_equivalence} derives the formulation of the design equations for the U-shaped springs using the derived stiffness matrices.

\subsection{Stiffness matrix ${[ {{\bf{K}}_{\rm{u}}} ]^{\rm{e}}}$ of U-shaped spring element} 
\label{sub:stiffness_matrix_of_u_shaped_spring_element}

For this derivation, the columns and crossbeam of the U-shaped spring are modeled as ideal Euler-Bernoulli beams, accounting solely for bending deformations. Shear deformation, axial deformation, and constrained torsion effects are neglected to simplify the analysis. The impact of shear deformation is explored separately in Section \ref{sub:numerical_validation_of_modal_frequencies}. Additionally, all deformations are assumed to be small, thereby excluding geometric nonlinearity.

\begin{figure}[H]
    \centering
\includegraphics[width=0.8\textwidth]{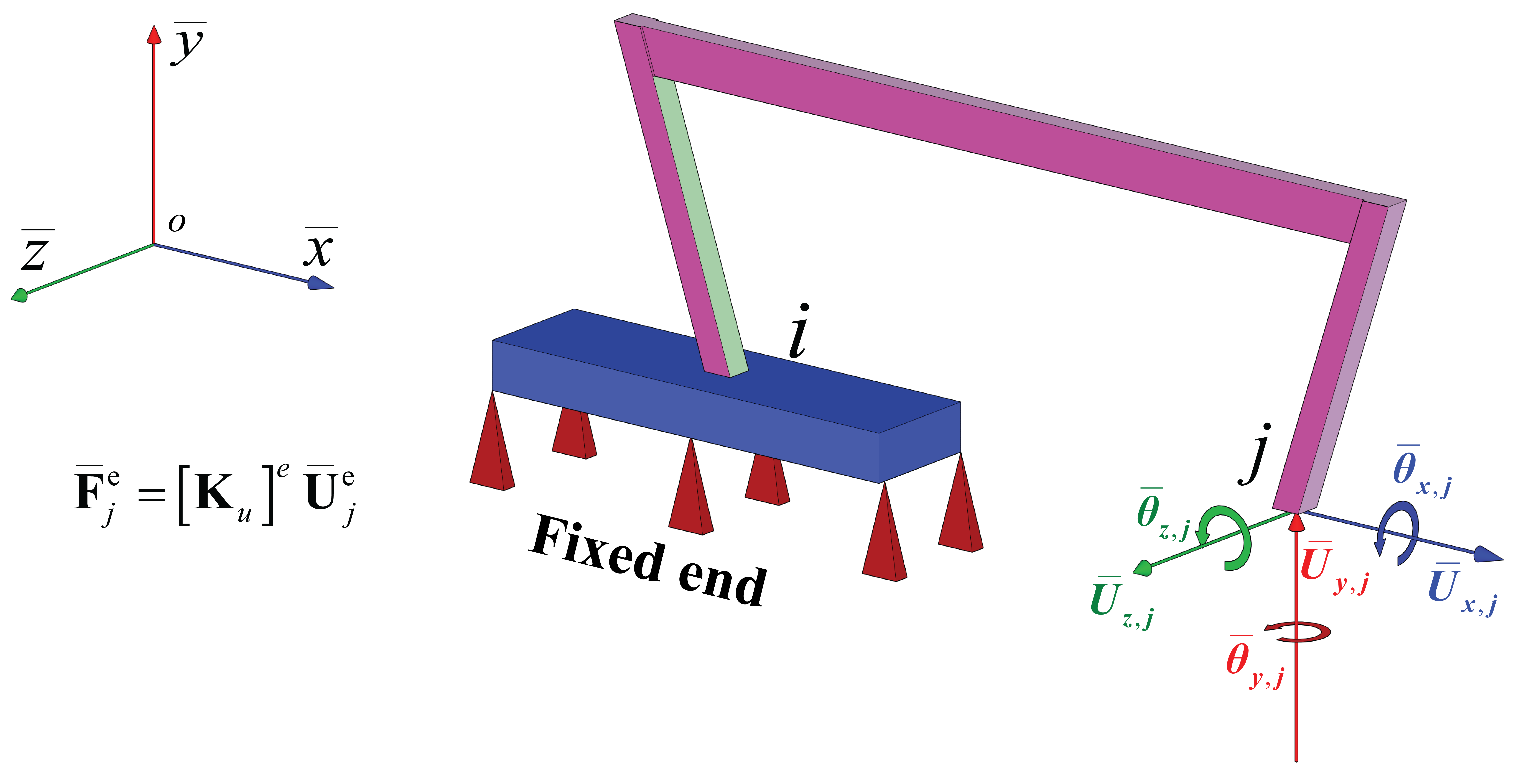}
    \caption{Schematic diagram illustrating the configuration of the U-shaped spring element.}
    \label{fig:7}
\end{figure}

For the cantilever U-shaped spring illustrated in Fig.~\ref{fig:7}, the relationship between nodal forces and displacements in the local coordinate system $\bar x\bar y\bar z$ is expressed through the stiffness matrix ${\left[ {{\bf{K}}_{\rm{u}}} \right]^{\rm{e}}}$. The detailed derivation of  ${\left[ {{\bf{K}}_{\rm{u}}} \right]^{\rm{e}}}$ follows in this subsection.
\begin{equation}
    \label{eq:12}
{\bf{\bar F}}_j^{\rm{e}} = {\left[ {{{\bf{K}}_{\rm{u}}}} \right]^{\rm{e}}}{\bf{\bar U}}_j^{\rm{e}}
\end{equation}
where ${\bf{\bar F}}_j^{\rm{e}}$ and ${\bf{\bar U}}_j^{\rm{e}}$ represent, respectively, the nodal force and displacement at the node $j$.
\begin{equation}
    \label{eq:13}
{\bf{\bar F}}_j^{\rm{e}} = {\left[ {\begin{array}{*{20}{c}}
{{{\bar F}_{x,j}}}&{{{\bar F}_{y,j}}}&{{{\bar F}_{z,j}}}&{{{\bar M}_{x,j}}}&{{{\bar M}_{y,j}}}&{{{\bar M}_{z,j}}}
\end{array}} \right]^{\rm{T}}}
\end{equation}
\begin{equation}
    \label{eq:14}
{\bf{\bar U}}_j^{\rm{e}} = {\left[ {\begin{array}{*{20}{c}}
{{{\bar U}_{x,j}}}&{{{\bar U}_{y,j}}}&{{{\bar U}_{z,j}}}&{{{\bar \theta }_{x,j}}}&{{{\bar \theta }_{y,j}}}&{{{\bar \theta }_{z,j}}}
\end{array}} \right]^{\rm{T}}}
\end{equation}

${\left[ {{\bf{K}}_{\rm{u}}} \right]^{\rm{e}}}$ denotes the stiffness matrix of a U-shaped spring element, which is expressed as
\begin{equation}
    \label{eq:15}
{\left[ {{{\bf{K}}_{\rm{u}}}} \right]^{\rm{e}}} = \left[ {\begin{array}{*{20}{c}}
{K_{11}^{\rm{e}}}&0&0&0&0&{K_{16}^{\rm{e}}}\\
0&{K_{22}^{\rm{e}}}&0&0&0&{K_{26}^{\rm{e}}}\\
0&0&{K_{33}^{\rm{e}}}&{K_{34}^{\rm{e}}}&{K_{35}^{\rm{e}}}&0\\
0&0&{K_{43}^{\rm{e}}}&{K_{44}^{\rm{e}}}&{K_{45}^{\rm{e}}}&0\\
0&0&{K_{53}^{\rm{e}}}&{K_{54}^{\rm{e}}}&{K_{55}^{\rm{e}}}&0\\
{K_{61}^{\rm{e}}}&{K_{62}^{\rm{e}}}&0&0&0&{K_{66}^{\rm{e}}}
\end{array}} \right]
\end{equation}

The stiffness matrix of a U-shaped spring element is symmetric according to the Betti's reciprocal‐work principle. It depends on the bending and torsional stiffness of its column and crossbeam, and can be expressed as
\begin{equation}
    \label{eq:16}
{\left[ {{{\bf{K}}_{\rm{u}}}} \right]^{\rm{e}}} = F\left( {{i_1},{i_2},{i_3},{i_4},{j_1},{j_2},{L_1},{L_2}} \right)
\end{equation}
where $i_1$, $i_2$ are, respectively, the in-plane bending stiffness of the column and crossbeam of the U-shaped spring
\begin{equation}
    \label{eq:17}
{i_1} = \frac{{En{m^3}}}{{12{L_1}}},{i_2} = \frac{{Ed{c^3}}}{{12{L_2}}}
\end{equation}
  $i_3$, $i_4$ are the out-of-plane bending stiffness of the column and crossbeam, respectively, which are expressed as
\begin{equation}
    \label{eq:18}
{i_3} = \frac{{Em{n^3}}}{{12{L_1}}},{i_4} = \frac{{Ec{d^3}}}{{12{L_2}}}
\end{equation}
  $j_1$, $j_2$ are the torsional stiffness of the column and crossbeam, respectively, which is defined as 
\begin{equation}
    \label{eq:19}
{j_1} = \frac{{G{J_1}}}{{{L_1}}},{j_2} = \frac{{G{J_2}}}{{{L_2}}}
\end{equation}
where $J_1$, $J_2$ are torsional constants, which can be evaluated for a solid rectangular section by using the following approximate formula with an error no greater than 4\% \citep{young2002roark}
\begin{equation}
    \label{eq:20}
J \approx \alpha {\beta ^3}\left[ {\frac{1}{3} - 0.21\frac{\beta }{\alpha }\left( {1 - \frac{{{\beta ^4}}}{{12{\alpha ^4}}}} \right)} \right],\;{\rm{for}}\;\alpha  \ge \beta \;
\end{equation}

In Eqs. (\ref{eq:18})-(\ref{eq:20}), the stiffness properties of U-shaped spring components are determined by their geometric parameters, which are illustrated in Fig. \ref{fig:3}\textit{a}. Specifically, for the calculation of ${j_1}$ via Eq. (\ref{eq:20}), the cross-sectional dimensions $m,n$ are first compared; the larger one is assigned to  $\alpha$ and the smaller one is assigned to $\beta$  in Eq.(\ref{eq:20}).

The detailed expression of ${\left[ {{{\bf{K}}_{\rm{u}}}} \right]^{\rm{e}}}$ is derived for a simplified scenario $L_2=L_3$. This derivation employs the unit displacement method \citep{gayed2021structural}, with the complete expression listed in Appendix A. For a more generalized case, where ${L_2} \ne {L_3}$, the stiffness matrix can be obtained through an analogous procedure. In this study, it is found that assuming $L_2=L_3$ makes the design of U-shaped springs sufficient for capturing the elastic stiffness properties. Consequently, the geometric vector $\boldsymbol{\rm{\upmu}}$ is reduced to ${[\begin{array}{*{20}{c}} m&n&c&d&{{L_1}}&{{L_2}} \end{array}]^{\rm{T}}}$.

The derivation of ${\left[ {{{\bf{K}}_{\rm{u}}}} \right]^{\rm{e}}}$ using the unit displacement method involves two key steps: (1) For each displacement component at the node $j$, apply a unit displacement while keeping all other displacement components fixed. (2) the resulting statically-indeterminate structure is analyzed to calculate the reaction forces at node  $j$. These reaction forces at node $j$ correspond to one column of the stiffness matrix  ${\left[ {{{\bf{K}}_{\rm{u}}}} \right]^{\rm{e}}}$. The condition of structural symmetry is exploited to simplify the analysis. The cantilever boundary condition, together with the symmetry of the U-shaped springs, enables a straightforward manual derivation of each component of  ${\left[ {{{\bf{K}}_{\rm{u}}}} \right]^{\rm{e}}}$.

From the derived components of ${\left[ {{{\bf{K}}_{\rm{u}}}} \right]^{\rm{e}}}$ in Appendix A, the following observations can be obtained: (1) the nodal displacement ${\bar U}_{x,j}$, ${\bar U}_{y,j}$  and ${\bar \theta}_{z,j}$ are associated with the in-plane bending of the U-shaped spring members. Accordingly, the corresponding stiffness terms ($K_{11}^{\rm{e}},K_{16}^{\rm{e}},K_{22}^{\rm{e}},K_{26}^{\rm{e}},K_{66}^{\rm{e}}$) depend exclusively on the in-plane bending stiffness $i_1,i_2$. (2) The nodal displacement ${\bar U}_{z,j}$, ${\bar \theta}_{x,j}$  and ${\bar \theta}_{y,j}$ are related to the out-of-plane bending and torsion. Thus, the associated stiffness terms ($K_{33}^{\rm{e}},K_{34}^{\rm{e}},K_{44}^{\rm{e}},K_{35}^{\rm{e}},K_{45}^{\rm{e}},K_{55}^{\rm{e}}$) are determined solely by the out-of-plane bending stiffness $i_3,i_4$ and torsional stiffness $j_1,j_2$ .

\subsection{Cantilever U-shaped spring element with a rigid beam} 
\label{sub:cantilever_u_shaped_spring_element_with_a_rigid_beam}

\begin{figure}[H]
    \centering
\includegraphics[width=0.95\textwidth]{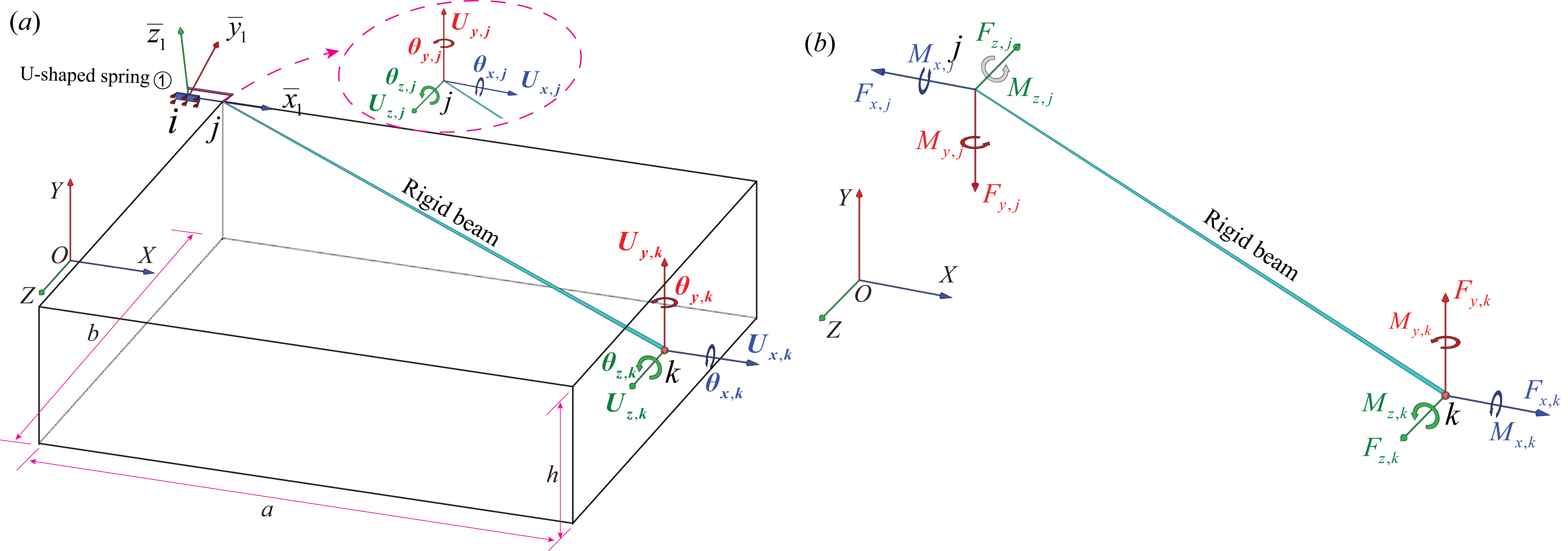}
    \caption{Schematic diagram of a cantilever U-shaped spring element connected with a rigid beam. (\textit{a}) The element coordinate system $\bar x_1 \bar y_1 \bar z_1$ and displacements of node $j,k$ in the global coordinate system  $XYZ$, (\textit{b}) Nodal forces of the rigid beam element.}
    \label{fig:8}
\end{figure}

Based on the stiffness matrix ${\left[ {{{\bf{K}}_{\rm{u}}}} \right]^{\rm{e}}}$ derived for the U-shaped spring element in the previous section, this section formulates the stiffness matrix for a cantilever U-shaped spring element connected to a rigid beam within the global coordinate system $XYZ$. The rigid beam represents the constraint imposed by the rigid frames. The relationship between the element's local coordinate system $\bar x_1 \bar y_1 \bar z_1$ and the global coordinate system $XYZ$ is illustrated in Fig. \ref{fig:8}.

In the global coordinate system $XYZ$, the rigid beam is described by a position vector:
\begin{equation}
    \label{eq:21}
{\left. {{\bf{kj}}} \right|_{XYZ}}{\rm{ = }} - a{{\bf{e}}_X} + {h_1}{{\bf{e}}_Y} - \frac{b}{2}{{\bf{e}}_Z}
\end{equation}
where ${\bf{e}}_X$, ${\bf{e}}_Y$, ${\bf{e}}_Z$ denote the unit vectors along the global $\mathbf{X,Y,Z}$ axes, respectively. $h_1$ represents the vertical distance between the shear center and anchor points of U-shaped springs at upper chord members, as shown in Fig. \ref{fig:9}\textit{b}.

Due to the constraint imposed by the rigid beam, a rotation at node $k$ (donated as $\theta_k$) induces a translational displacement at node $j$. Assuming small deformations, this kinematic relationship is expressed as:
\begin{equation}
    \label{eq:22}
{U_{x,j}} = {U_{x,k}} - \frac{b}{2}{\theta _{y,k}} - {h_1}{\theta _{z,k}}
\end{equation}
\begin{equation}
    \label{eq:23}
{U_{y,j}} = {U_{y,k}} + \frac{b}{2}{\theta _{x,k}} - a{\theta _{z,k}}
\end{equation}
\begin{equation}
    \label{eq:24}
{U_{z,j}} = {U_{z,k}} + {h_1}{\theta _{x,k}} + a{\theta _{y,k}}
\end{equation}

Due to the rigid body movement, the torsional angles of node $k$ and node $j$  remain identical
\begin{equation}
    \label{eq:25}
{\theta _{x,j}} = {\theta _{x,k}},{\theta _{y,j}} = {\theta _{y,k}},{\theta _{z,j}} = {\theta _{z,k}}
\end{equation}

The relationship in Eqs. (\ref{eq:22})-(\ref{eq:25}) can be rearranged into a matrix form as
\begin{equation}
    \label{eq:26}
{{\bf{U}}_j} = {{\bf{T}}_1}{{\bf{U}}_k}
\end{equation}
where ${\bf{U}}_j$, ${\bf{U}}_k$ are the nodal displacement vectors at node $j$ and node $k$, which are expressed as
\begin{equation}
    \label{eq:27}
{{\bf{U}}_j} = {\left[ {\begin{array}{*{20}{c}}
{{U_{x,j}}}&{{U_{y,j}}}&{{U_{z,j}}}&{{\theta _{x,j}}}&{{\theta _{y,j}}}&{{\theta _{z,j}}}
\end{array}} \right]^{\mathop{\rm T}\nolimits} }
\end{equation}
\begin{equation}
    \label{eq:28}
{{\bf{U}}_k} = {\left[ {\begin{array}{*{20}{c}}
{{U_{x,k}}}&{{U_{y,k}}}&{{U_{z,k}}}&{{\theta _{x,k}}}&{{\theta _{y,k}}}&{{\theta _{z,k}}}
\end{array}} \right]^{\mathop{\rm T}\nolimits} }
\end{equation}
\begin{equation}
    \label{eq:29}
{{\bf{T}}_1} = \left[ {\begin{array}{*{20}{c}}
1&0&0&{ - \frac{b}{2}}&{ - {h_1}}&0\\
0&1&0&{\frac{b}{2}}&0&{ - a}\\
0&0&1&{{h_1}}&a&0\\
0&0&0&1&0&0\\
0&0&0&0&1&0\\
0&0&0&0&0&1
\end{array}} \right]
\end{equation}

In Fig. \ref{fig:8}\textit{b}, the nodal forces at node $j$ of the rigid beam are the reciprocal forces to the U-shaped spring element's nodal forces at this node; Hence, these force components are in the negative direction. The forces at node $j$  induce moments at node $k$, the induced moments can be generally expressed as $\Delta {{\bf{M}}_k} = {\left. {{\bf{kj}}} \right|_{XYZ}} \times \left( {{F_{x,j}}{{\bf{e}}_X} + {F_{y,j}}{{\bf{e}}_Y} + {F_{z,j}}{{\bf{e}}_Z}} \right) $ , which can also be derived from the equilibrium condition of the rigid beam and expressed as
\begin{equation}
    \label{eq:30}
{M_{x,k}} = {M_{x,j}} + {h_1}{F_{z,j}}{\rm{ + }}\frac{b}{2}{F_{y,j}}
\end{equation}
\begin{equation}
    \label{eq:31}
{M_{y,k}} = {M_{y,j}} + a{F_{z,j}} - \frac{b}{2}{F_{x,j}}
\end{equation}
\begin{equation}
    \label{eq:32}
{M_{z,k}} = {M_{z,j}} - a{F_{y,j}} - {h_1}{F_{x,j}}
\end{equation}

Each force component at node $j$ and node $k$ should be identical to satisfy the equilibrium condition, that is
\begin{equation}
    \label{eq:33}
{F_{x,j}} = {F_{x,k}},{F_{y,j}} = {F_{y,k}},{F_{z,j}} = {F_{z,k}}
\end{equation}

The above relationship can be expressed into 
\begin{equation}
    \label{eq:34}
{{\bf{F}}_k} = {\bf{T}}_1^{\rm{T}}{{\bf{F}}_j}
\end{equation}
where
\begin{equation}
    \label{eq:35}
{{\bf{F}}_j} = {\left[ {\begin{array}{*{20}{c}}
{{F_{x,j}}}&{{F_{y,j}}}&{{F_{z,j}}}&{{M_{x,j}}}&{{M_{y,j}}}&{{M_{z,j}}}
\end{array}} \right]^{\mathop{\rm T}\nolimits} }
\end{equation}
\begin{equation}
    \label{eq:36}
{{\bf{F}}_k} = {\left[ {\begin{array}{*{20}{c}}
{{F_{x,k}}}&{{F_{y,k}}}&{{F_{z,k}}}&{{M_{x,k}}}&{{M_{y,k}}}&{{M_{z,k}}}
\end{array}} \right]^{\mathop{\rm T}\nolimits} }
\end{equation}

The nodal displacement vector of node $j$ in the the global coordinate system  $XYZ$ can be further transformed into the local coordinate system $\bar x_1 \bar y_1 \bar z_1$ using the rotation matrix as
\begin{equation}
    \label{eq:37}
{\bf{\bar U}}_j^{\rm{e}} = {{\bf{\lambda }}_1}{{\bf{U}}_j}
\end{equation}
where $\lambda_1$ is the rotation matrix and expressed as
\begin{equation}
    \label{eq:38}
{{\bf{\lambda }}_1} = \left[ {\begin{array}{*{20}{c}}
{{{\bf{e}}_X} \cdot {{\bf{e}}_{{x_1}}}}&{{{\bf{e}}_Y} \cdot {{\bf{e}}_{{x_1}}}}&{{{\bf{e}}_Z} \cdot {{\bf{e}}_{{x_1}}}}\\
{{{\bf{e}}_X} \cdot {{\bf{e}}_{{y_1}}}}&{{{\bf{e}}_Y} \cdot {{\bf{e}}_{{y_1}}}}&{{{\bf{e}}_Z} \cdot {{\bf{e}}_{{y_1}}}}\\
{{{\bf{e}}_X} \cdot {{\bf{e}}_{{z_1}}}}&{{{\bf{e}}_Y} \cdot {{\bf{e}}_{{z_1}}}}&{{{\bf{e}}_Z} \cdot {{\bf{e}}_{{z_1}}}}
\end{array}} \right] = \left[ {\begin{array}{*{20}{c}}
1&0&0\\
0&{\cos {\theta _1}}&{ - \sin {\theta _1}}\\
0&{\sin {\theta _1}}&{\cos {\theta _1}}
\end{array}} \right]
\end{equation}

Similarly, the transformation of nodal force vector at node $j$ from the local coordinate system $\bar x_1 \bar y_1 \bar z_1$ to the global coordinate system $XYZ$ as
\begin{equation}
    \label{eq:39}
{{\bf{F}}_j} = {\bf{\lambda }}_1^{\rm{T}}{\bf{\bar F}}_j^{\rm{e}}
\end{equation}

Substituting Eqs. (\ref{eq:12}), (\ref{eq:26}), (\ref{eq:37}), (\ref{eq:39}) into Eq. (\ref{eq:34}), yields
\begin{equation}
    \label{eq:40}
\begin{array}{l}
{{\bf{F}}_k}\xrightarrow{\text{ Eq. (34) }}{\bf{T}}_1^{\rm{T}}{{\bf{F}}_j}\xrightarrow{\text{ Eq. (39) }}{\bf{T}}_1^{\rm{T}}{\bf{\lambda }}_1^{\rm{T}}{\bf{\bar F}}_j^{\rm{e}}\xrightarrow{\text{ Eq. (12) }}{\bf{T}}_1^{\rm{T}}{\bf{\lambda }}_1^{\rm{T}}{\left[ {{{\bf{K}}_u}} \right]^{\rm{e}}}{\bf{\bar U}}_j^{\rm{e}}\\
\;\;\;\;\;\xrightarrow{\text{ Eq. (37) }}{\bf{T}}_1^{\rm{T}}{\bf{\lambda }}_1^{\rm{T}}{\left[ {{{\bf{K}}_u}} \right]^{\rm{e}}}{{\bf{\lambda }}_1}{{\bf{U}}_j}\xrightarrow{\text{ Eq. (26) }}\left( {{\bf{T}}_1^{\rm{T}}{\bf{\lambda }}_1^{\rm{T}}{{\left[ {{{\bf{K}}_u}} \right]}^{\rm{e}}}{{\bf{\lambda }}_1}{{\bf{T}}_1}} \right){{\bf{U}}_k}
\end{array}
\end{equation}

Eq. (\ref{eq:40}) is simplified as 
\begin{equation}
    \label{eq:41}
{{\bf{F}}_k} = {\left[ {{{\bf{K}}_1}} \right]^{\rm{e}}}{{\bf{U}}_k}
\end{equation}
where ${\left[ {{{\bf{K}}_{1}}} \right]^{\rm{e}}}$ is the stiffness matrix of the cantilever U-shaped spring connected to a rigid beam in the global coordinate system (see Fig. \ref{fig:8}\textit{a}), and it is obtained by transforming the local stiffness matrix ${\left[ {{{\bf{K}}_{\rm{u}}}} \right]^{\rm{e}}}$ of a U-shaped spring element (derived in the previous section) via the rotation matrix and by imposing the rigid-beam constraint. Inferring from Eq. (\ref{eq:40}), ${\left[ {{{\bf{K}}_1}} \right]^{\rm{e}}}$ is expressed as
\begin{equation}
    \label{eq:42}
{\left[ {{{\bf{K}}_1}} \right]^{\rm{e}}} = {\bf{T}}_1^{\rm{T}}{\bf{\lambda }}_1^{\rm{T}}{\left[ {{{\bf{K}}_u}} \right]^{\rm{e}}}{{\bf{\lambda }}_1}{{\bf{T}}_1}
\end{equation}

It is found from Eq. (\ref{eq:42}) that just like ${\left[ {{{\bf{K}}_{\rm{u}}}} \right]^{\rm{e}}}$, ${\left[ {{{\bf{K}}_1}} \right]^{\rm{e}}}$ is also a real-valued symmetric matrix.

\subsection{Stiffness matrix ${[\bf{K}]^{\rm{e}}}$ of a cantilever spring-lever element} 
\label{sub:stiffness_matrix_of_a_cantilever_spring_lever_element}

\begin{figure}[H]
    \centering
\includegraphics[width=0.8\textwidth]{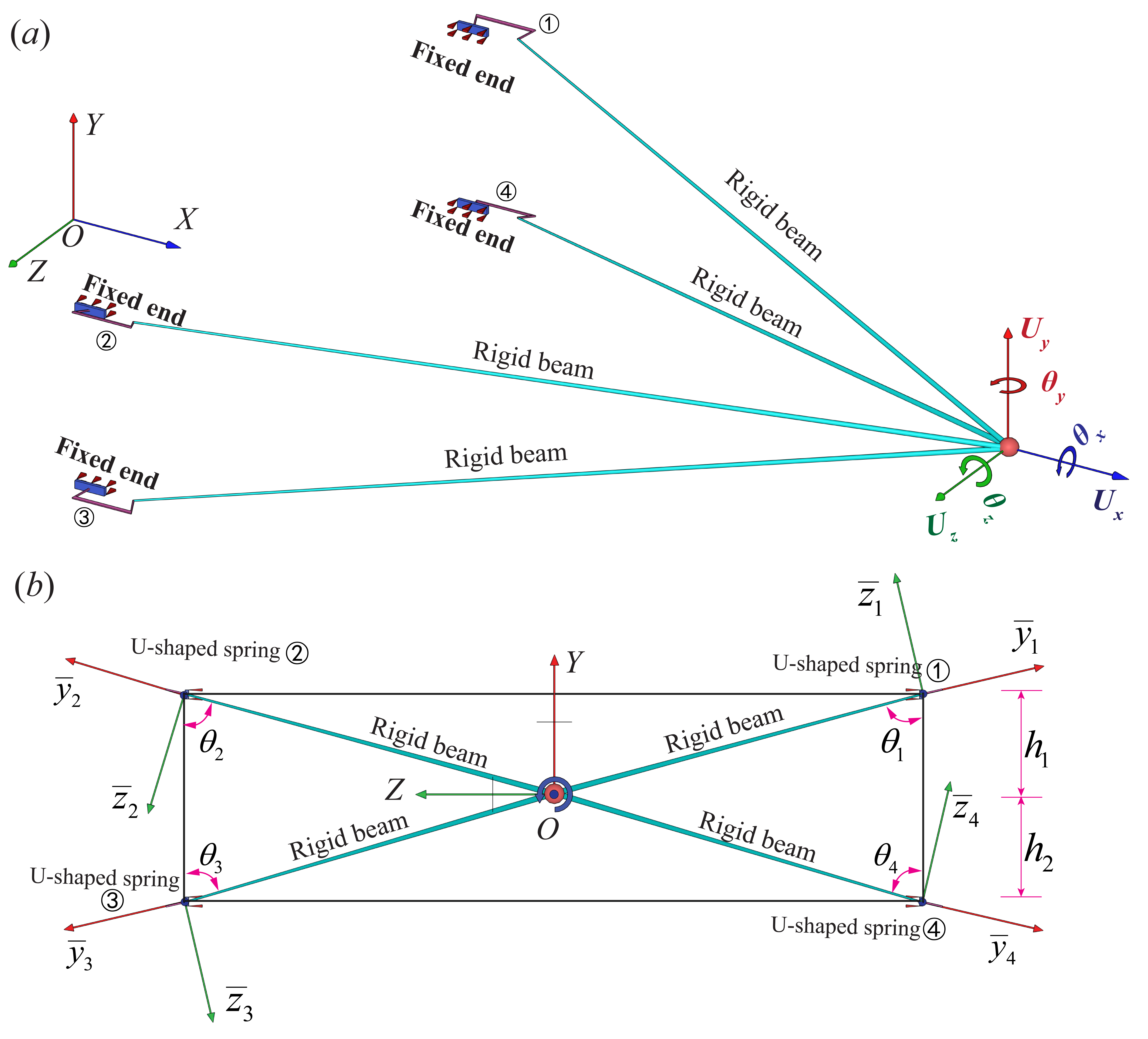}
    \caption{The spring-lever element consisting of four U-shaped spring elements and rigid beams. (\textit{a}) Axonometric View,  (\textit{b}) View along the negative X-axis direction of the global cartesian coordinate system $XYZ$.}
    \label{fig:9}
\end{figure}

The spring-lever element in Fig. \ref{fig:9} is a simplified mechanical analog of the cantilever rigid-frame element in Fig. \ref{fig:3}\textit{b}, where the constraint of the rigid frame on the four U-shaped springs is simplified as four rigid beams. The U-shaped springs are numbered as \ding{172}, \ding{173}, \ding{174}, \ding{175} and the corresponding local coordinates are denoted as ${\bar x_i}{\bar y_i}{\bar z_i}\left( {i = 1,2,3,4} \right)$ in Fig. \ref{fig:9}. The derived stiffness matrix ${\left[ {{{\bf{K}}_1}} \right]^{\rm{e}}}$ in Section \ref{sub:cantilever_u_shaped_spring_element_with_a_rigid_beam} corresponds to the U-shaped spring numbered as \ding{172}.

As the four U-shaped springs are connected in series, the stiffness matrix of their assembly in the spring-lever element can be formulated as their combination in the global coordinate system, which is expressed as 
\begin{equation}
    \label{eq:43}
{[{\bf{K}}]^{\rm{e}}} = \sum\limits_{i = 1}^4 {{{[{{\bf{K}}_i}]}^{\rm{e}}}}  = \sum\limits_{i = 1}^4 {{{\left( {{{\bf{\lambda }}_i}{{\bf{T}}_i}} \right)}^{\rm{T}}}{{\left[ {{{\bf{K}}_\text{u}}} \right]}^{\rm{e}}}\left( {{{\bf{\lambda }}_i}{{\bf{T}}_i}} \right)}
\end{equation}
where $\lambda_i$ denotes the rotation matrix from the global coordinate system $XYZ$ to the element coordinate system ${\bar x_i}{\bar y_i}{\bar z_i}$. $\lambda_1$ is expressed in Eq. (\ref{eq:38}) and others are expressed as 
\begin{subequations}
\label{eq:44}
\begin{align}
\mathbf{\lambda}_2 &= \begin{bmatrix}
1 & 0 & 0 \\
0 & \cos \theta_2 & \sin \theta_2 \\
0 & -\sin \theta_2 & \cos \theta_2
\end{bmatrix}, \tag{44a} \\
\mathbf{\lambda}_3 &= \begin{bmatrix}
1 & 0 & 0 \\
0 & -\cos \theta_3 & \sin \theta_3 \\
0 & -\sin \theta_3 & -\cos \theta_3
\end{bmatrix}, \tag{44b} \\
\mathbf{\lambda}_4 &= \begin{bmatrix}
1 & 0 & 0 \\
0 & -\cos \theta_4 & -\sin \theta_4 \\
0 & \sin \theta_4 & -\cos \theta_4
\end{bmatrix}. \tag{44c}
\end{align}
\end{subequations}

${{\bf{T}}_i}$ represents the constraint of nodal displacements and forces by the rigid beams, which can be derived following a similar procedure in Section \ref{sub:cantilever_u_shaped_spring_element_with_a_rigid_beam}. ${{\bf{T}}_1}$ is expressed in Eq. (\ref{eq:29}) and others can be easily transformed from ${{\bf{T}}_1}$ by replacing the relevant geometric parameters as
\begin{equation}
    \label{eq:45}
{{\bf{T}}_{\rm{2}}} = {\left. {{{\bf{T}}_1}} \right|_{b \to  - b}},{{\bf{T}}_3} = {\left. {{{\bf{T}}_1}} \right|_{b \to  - b,{h_1} \to  - {h_2}}},{{\bf{T}}_4} = {\left. {{{\bf{T}}_1}} \right|_{{h_1} \to  - {h_2}}}
\end{equation}
where $h_2$ is defined in Fig. \ref{fig:9}\textit{b}, denoting the vertical distance between the shear center and anchor points of U-shaped springs at lower chord members.

\subsection{Design equations by elastic-strain-energy equivalence} 
\label{sub:design_equations_by_elastic_strain_energy_equivalence}

Intuitively, the design equation for the geometric parameter $\boldsymbol{\upmu}$  is directly obtained by matching each term of the stiffness matrix ${\left[ \bf{K} \right]^{\rm{e}}}$, derived in Section \ref{sub:stiffness_matrix_of_a_cantilever_spring_lever_element}, with that of a cantilever Euler-Bernoulli beam, as shown in Fig. \ref{fig:4}. However, a direct, term-by-term identification, fails to yield a solution, since the cross-stiffness terms of a spring-lever element behave differently with an Euler-Bernoulli beam.

The stiffness matrix of a cantilever Euler-Bernoulli beam can be expressed as
\begin{equation}
    \label{eq:46}
\left[ {\bf{K}} \right]_{{\rm{el}}}^{\rm{e}} = \left[ {\begin{array}{*{20}{c}}
{\frac{{EA}}{l}}&0&0&0&0&0\\
0&{\frac{{12E{I_z}}}{{{l^3}}}}&0&0&0&{\frac{{6E{I_z}}}{{{l^2}}}}\\
0&0&{\frac{{12E{I_y}}}{{{l^3}}}}&0&{ - \frac{{6E{I_y}}}{{{l^2}}}}&0\\
0&0&0&{\frac{{GJ}}{l}}&0&0\\
0&0&{ - \frac{{6E{I_y}}}{{{l^2}}}}&0&{\frac{{4E{I_y}}}{l}}&0\\
0&{\frac{{6E{I_z}}}{{{l^2}}}}&0&0&0&{\frac{{4E{I_z}}}{l}}
\end{array}} \right]
\end{equation}

It is observed from Eq. (\ref{eq:46}) that the off-diagonal terms exhibit specific proportional relationships with their corresponding diagonal elements, for instance ${\bf{K}}_{{\rm{el}}}^{\rm{e}}\left( {2,2} \right)/{\bf{K}}_{{\rm{el}}}^{\rm{e}}\left( {2,6} \right) = 2/l$, ${\bf{K}}_{{\rm{el}}}^{\rm{e}}\left( {5,3} \right)/{\bf{K}}_{{\rm{el}}}^{\rm{e}}\left( {5,5} \right) = -1.5/l$. These quantitive relationships stem from the shape functions and moment-curvature equations governing an Euler-Bernoulli beam. However, these relationships do not apply to the spring-lever element, as the rigid beams impose constraints without undergoing deformation. 

In the following section, the design equations are established via the principle of elastic-strain energy equivalence. This approach ensures that the strain energy of a cantilever beam, consisting of spring-lever elements, matches that of an equivalent Euler-Bernoulli beam. By achieving this energy equivalence, the discrete system can effectively approximate the girder's strain energy and deflection behavior of a truss girder using a finite number of discrete segments. As the number of spring-lever elements increases, the approximation approaches exact equivalence, making this method particularly suitable for long-span truss girders.

\subsubsection{Equivalence in bending strain energy} 
\label{ssub:equivalence_in_bending_strain_energy}

For a cantilever Euler-Bernoulli beam subjected to a load applied at its free end in positive $Y$-direction, the elastic-strain-energy stored in the beam equals the work done by the external work. This relationship is expressed as
\begin{equation}
    \label{eq:47}
{U_{{\rm{el}}}} = \frac{1}{2}F{v_y} = \frac{{{F^2}{n^3}{l^3}}}{{6E{I_z}}}
\end{equation}
where $v_y$ is the vertical deflection at the free end and calculated by Eq. (\ref{eq:6}), where the span length is $nl$ with $n$ being the number of discrete segments and $l$ being the axial length of one discrete segment. 

\begin{figure}[H]
    \centering
\includegraphics[width=0.8\textwidth]{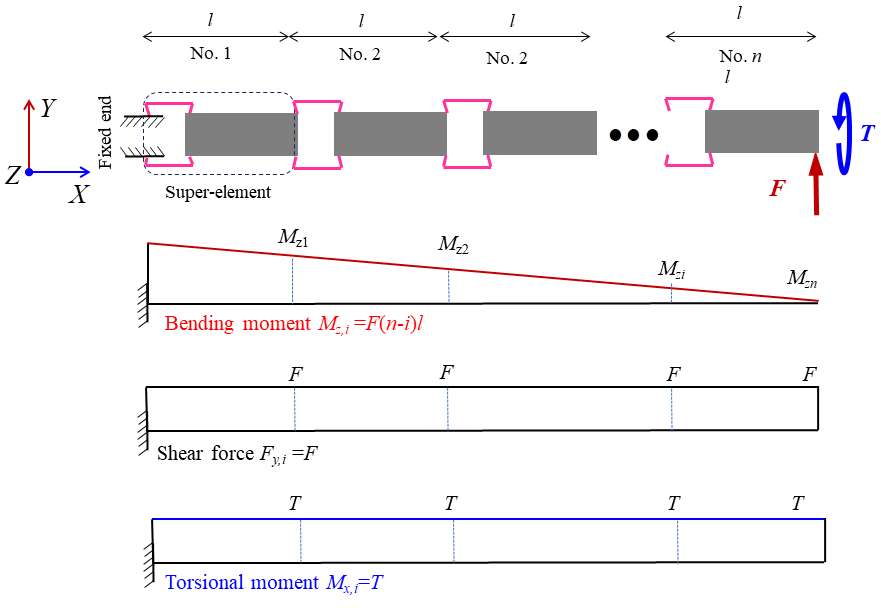}
    \caption{Schematic diagram of a cantilever beam consisting of spring-lever elements connected end-to-end and the corresponding internal forces.}
    \label{fig:10}
\end{figure}

As shown in Fig. \ref{fig:10}, the total elastic strain-energy of the discrete stiffness system equals the sum of work done at each cantilever element, which are loaded at their free end by the internal bending moment $M_{z,i}$ and shear force $F_{y,i}$. The distribution of  $M_{z,i}$ and $F_{y,i}$ are depicted in Fig. \ref{fig:10}. The sum of elastic-strain-energy is expressed as
\begin{equation}
\label{eq:48}
\begin{aligned}
U_{\text{el}} &= \sum_{i=1}^n \frac{1}{2} \begin{bmatrix} F_{y,i} & M_{z,i} \end{bmatrix} \begin{bmatrix} U_{y,i} & \theta_{z,i} \end{bmatrix}^{\text{T}} \\
&= \frac{1}{2} F^2 \sum_{i=1}^n \begin{bmatrix} 1 & l(n - i) \end{bmatrix} \left[ \mathbf{K} \right]_{26}^{-1} \begin{bmatrix} 1 & l(n - i) \end{bmatrix}^{\text{T}}.
\end{aligned}
\end{equation}
\begin{equation}
    \label{eq:49}
\left[ {\bf{K}} \right]_{26}^{} = \left[ {\begin{array}{*{20}{c}}
{{{\bf{K}}^{\rm{e}}}\left( {2,2} \right)}&{{{\bf{K}}^{\rm{e}}}\left( {2,6} \right)}\\
{{{\bf{K}}^{\rm{e}}}\left( {6,2} \right)}&{{{\bf{K}}^{\rm{e}}}\left( {6,6} \right)}
\end{array}} \right]
\end{equation} 

The energy equivalence between the discrete stiffness system and a Euler-Bernoulli beam leads to
\begin{equation}
    \label{eq:50}
\sum\limits_{i = 1}^n {\left[ {\begin{array}{*{20}{c}}
1&{l\left( {n - i} \right)}
\end{array}} \right]} \left[ {\bf{K}} \right]_{26}^{ - 1}{\left[ {\begin{array}{*{20}{c}}
1&{l\left( {n - i} \right)}
\end{array}} \right]^{\rm{T}}} = \frac{{{n^3}{l^3}}}{{3E{I_z}}}
\end{equation} 

Eq. (\ref{eq:48}) represents the design equation for vertical bending in an implicit form. If the elastic strain due to shear force is neglected and only the strain induced by the bending moment is considered---since at first glance, the calculation in Eq. (\ref{eq:47}) accounts solely for bending deformation---an explicit form can be obtained as follows:
\begin{equation}
    \label{eq:51}
{K_{66}} = {{\bf{K}}^{\rm{e}}}\left( {6,6} \right) = \frac{{3\sum\limits_{i = 1}^n {{{\left( {n - i} \right)}^2}} }}{{{n^3}}}\frac{{E{I_z}}}{l}
\end{equation}

It is beneficial to further compare the explicit form with a hinged-support beam. As shown in Fig. \ref{fig:10}, a hinged-support beam is applied by a uniformly distributed load $q$. The elastic-strain energy for the corresponding Euler-Bernoulli beam equals to the external work by $q$, which is expressed as
\begin{equation}
    \label{eq:52}
{U_{{\rm{el}}}} = \frac{1}{2}\int_0^L {E{I_z}{v''_y}}  \cdot {v''_y}{\mathop{\rm d}\nolimits} x = \int_0^{L/2} {\frac{{M_z^2\left( x \right)}}{{E{I_z}}}} {\mathop{\rm d}\nolimits} x = \frac{{{q^2}{n^5}{l^5}}}{{{\rm{240}}E{I_z}}}
\end{equation}

The elastic-strain energy for a discrete stiffness system is expressed in a similar form of Eq. (\ref{eq:50}), which is further approximated by neglecting the shear terms as
\begin{equation}
    \label{eq:53}
{U_{{\rm{el}}}} = \sum\limits_{i = 1}^n {\left( {\frac{1}{2}{M_{z,i}}{\theta _{z,i}}} \right)}  = \frac{1}{{{\rm{2}}{K_{66}}}}\sum\limits_{i = 1}^n {M_{z,i}^2}  = \frac{1}{{4{K_{66}}}}\sum\limits_{i = 1}^{n{\rm{/2}}} {{q^2}{i^2}{l^4}{{\left( {n - i} \right)}^2}}
\end{equation}

Equating Eq. (\ref{eq:52}) and Eq. (\ref{eq:53}), one has
\begin{equation}
    \label{eq:54}
{K_{66}} = \frac{{{\rm{60}}\sum\limits_{i = 1}^{n{\rm{/2}}} {{i^2}{{\left( {n - i} \right)}^2}} }}{{{n^5}}}\frac{{E{I_z}}}{l}
\end{equation}

\begin{figure}[H]
    \centering
\includegraphics[width=0.8\textwidth]{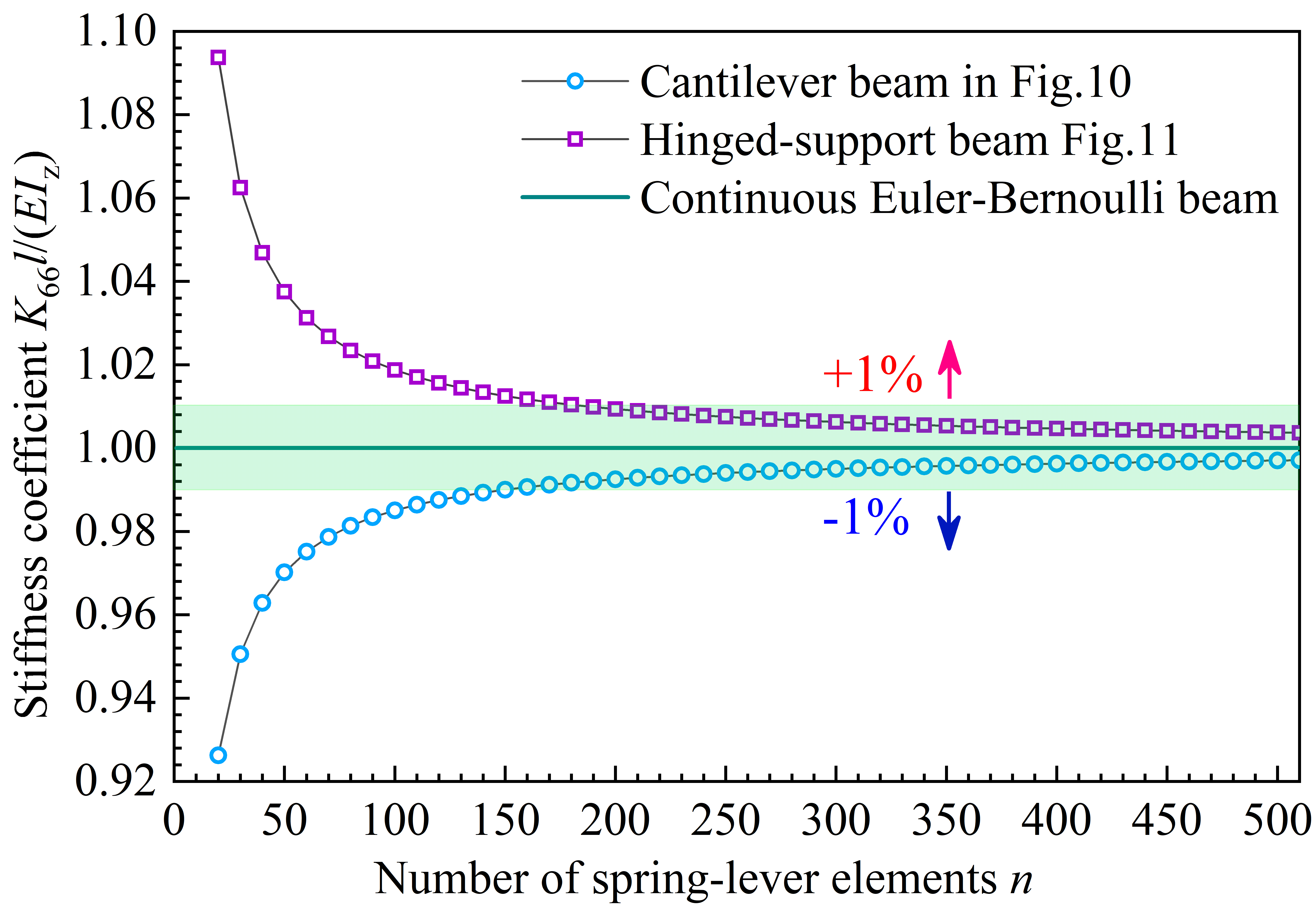}
    \caption{The equivalent bending stiffness $K_{66}$ with the number of spring-lever elements, where the discrete-stiffness cantilever beam in Eq. (\ref{eq:51}), and discrete-stiffness hinged-support beam in Eq. (\ref{eq:54}), and the continuous-stiffness Euler-Bernoulli beam are compared.}
    \label{fig:12}
\end{figure}

The stiffness term $K_{66}$ calculated for spring-lever elements using Eq. (\ref{eq:51}) and Eq. (\ref{eq:54}) is compared with that of an equivalent Euler-Bernoulli beam. The obtained stiffness $K_{66}$ is non-dimensionalized with respect to the Euler-Bernoulli beam stiffness ${{E{I_z}}}/{l}$ and is presented in Fig. \ref{fig:12}. The analysis reveals that the stiffness $K_{66}$ of a system with a finite number of spring-lever elements deviates from that of the Euler-Bernoulli beam required for strain energy equivalence, primarily due to differences in deformation patterns. The influence of supporting conditions is significant when the number of segments $n$ is less than 50. However, as $n$ increases, the discrete system converges toward the behavior of a continuous-stiffness Euler-Bernoulli beam, with the discrepancy reducing to less than $\pm 1\%$ when $n$ exceeds 170.

\begin{figure}[H]
    \centering
\includegraphics[width=0.8\textwidth]{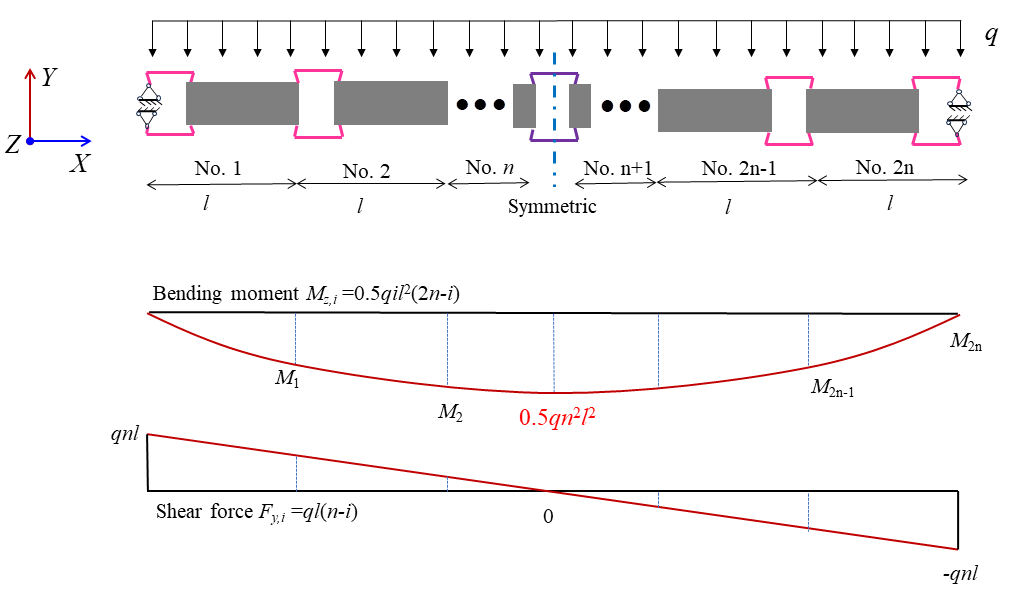}
    \caption{A hinged-support beam consisting of spring-lever elements connected end-to-end and the corresponding internal forces.}
    \label{fig:11}
\end{figure}

The variation in supporting conditions stems from differences in the distributions of bending moment $M_{z,i}$ and shear force $F_{y,i}$. This effect can be mitigated by incorporating shear strain, which, although neglected in the Euler-Bernoulli beam model, contributes additional strain energy in the spring-lever elements. Specifically, the shear force at node $k$ induces a bending moment at node $j$, as illustrated in Fig. \ref{fig:8}. Additionally, for the hinged-support beam depicted in Fig. \ref{fig:11}, the elastic strain of U-shaped springs at the midspan is neglected, as these cannot be combined with a rigid block to form a cantilever spring-lever element. Consequently, the discrepancy in the hinged-support case is greater than that observed in the cantilever configuration shown in Fig. \ref{fig:12}. Therefore, the implicit form of the design equation, Eq. (\ref{eq:48}), is recommended when the number of spring-lever elements $n$is no larger than 170.

Following a similar methodology, the design equation for transverse bending is derived as follows:
\begin{equation}
    \label{eq:55}
\sum\limits_{i = 1}^n {\left[ {\begin{array}{*{20}{c}}
1&{l\left( {n - i} \right)}
\end{array}} \right]} \left[ {\bf{K}} \right]_{{\rm{35}}}^{ - 1}{\left[ {\begin{array}{*{20}{c}}
1&{l\left( {n - i} \right)}
\end{array}} \right]^{\rm{T}}} = \frac{{{n^3}{l^3}}}{{3E{I_y}}}
\end{equation}
where the matrix $\left[ {\bf{K}} \right]_{{\rm{35}}}$ is expressed as
\begin{equation}
    \label{eq:56}
\left[ {\bf{K}} \right]_{35}^{} = \left[ {\begin{array}{*{20}{c}}
{{{\bf{K}}^{\rm{e}}}\left( {3,3} \right)}&{{{\bf{K}}^{\rm{e}}}\left( {3,5} \right)}\\
{{{\bf{K}}^{\rm{e}}}\left( {5,3} \right)}&{{{\bf{K}}^{\rm{e}}}\left( {5,5} \right)}
\end{array}} \right]
\end{equation}


\subsubsection{Equivalence in torsional strain energy} 
\label{ssub:equivalence_in_torsional_strain_energy}

The design equation for the torsional stiffness of the discrete-stiffness system can be obtained analogously. For a cantilever Euler-Bernoulli beam subjected to a torque $T$ applied at its free end, the elastic strain energy stored in the beam equals the work performed by the torque, expressed as:
\begin{equation}
    \label{eq:57}
{U_{{\rm{el}}}} = \frac{1}{2}T{\theta _x} = \frac{{{T^2}nl}}{{2G{J_{\rm{d}}}}}
\end{equation}
where $\theta _x$ represents the torsional angle at the free end.  

In the cantilever discrete stiffness system, as illustrated in Fig. \ref{fig:10}, the total elastic strain energy accumulated during torsional deformation is calculated as the sum of the work done by the internal torsional moments acting on each cantilever element. This is formulated as:
\begin{equation}
    \label{eq:58}
{U_{{\rm{el}}}} = \sum\limits_{i = 1}^n {\frac{1}{2}{M_{x,i}}{\theta _{x,i}}}  = \frac{1}{2}T\sum\limits_{i = 1}^n {{\theta _{x,i}}}  = \frac{{{T^2}n}}{{2{{\bf{K}}^{\rm{e}}}\left( {4,4} \right)}}
\end{equation}

The equivalence of Eq. (\ref{eq:57}) and Eq. (\ref{eq:58}) leads to
\begin{equation}
    \label{eq:59}
{{\bf{K}}^{\rm{e}}}\left( {4,4} \right) = \frac{{G{J_{\rm{d}}}}}{l}
\end{equation}

Eqs. (\ref{eq:48}), (\ref{eq:55}), and (\ref{eq:58}) constitute the closed-form design equations for the U-shaped springs within the discrete stiffness system. These equations provide the explicit functional form of Eq. (\ref{eq:4}), linking the geometric parameters of the U-shaped springs to the bending and torsional stiffness of the truss girder. The derivation demonstrates that any set of geometric parameters satisfying these equations ensures that the truss girder exhibits strain energy equivalent to that of the corresponding Euler-Bernoulli beam.


\section{Optimization of design parameters and numerical validation} 
\label{sec:optimization_of_design_parameters_and_numerical_validation}

The geometric parameters of the U-shaped springs are determined by solving the design equations derived in Section \ref{sec:derivation_of_design_equations_for_the_u_shaped_springs} through a global optimization procedure, detailed in Section \ref{sub:optimization_algorithm}. The feasibility and accuracy of these calculated parameters are validated numerically in Section \ref{sub:numerical_validation_of_modal_frequencies}, focusing on modal frequencies, and in Section \ref{sub:comparison_of_aerostatic_deformation}, examining aerostatic deformation.

\subsection{Optimization algorithm} 
\label{sub:optimization_algorithm}

As outlined in Section \ref{sub:discrete_stiffness_system_and_design_procedure_of_u_shaped_springs}, the design of U-shaped springs requires the discrete stiffness system to replicate the elastic properties of the truss girder, as governed by the design equations in Section \ref{sub:design_equations_by_elastic_strain_energy_equivalence}. This design task is formulated as a constrained optimization problem:
\begin{equation}
    \label{eq:60}
\left\{ \begin{array}{l}
\mathop {\min }\limits_{\boldsymbol{\upmu }} \;J\left( {\boldsymbol{\upmu }} \right) = {w_1}{r_z}^2\left( {\boldsymbol{\upmu }} \right) + {w_2}{r_y}^2\left( {\boldsymbol{\upmu }} \right) + {w_3}{r_x}^2\left( {\boldsymbol{\upmu }} \right)\\
{\rm{subject}}\;{\rm{to}}\;{\boldsymbol{\upmu }} = 0.1 \times {\bf{k}},\;{\bf{k}} \in {\mathbb{Z}^6},\;{k_{i,\min }} \le {k_i} \le {k_{i,\max }},\;i = 1,2, \ldots ,6
\end{array} \right.
\end{equation}
where the objective function $J$ represents the weighted sum of residuals from the design equations. $r_z,r_y,r_x$ denote, respectively, the residuals of Eqs. (\ref{eq:48}), (\ref{eq:53}), and (\ref{eq:57}). ${w_i}\left( {i = 1,2,3} \right)$ represents their respective weights. $\boldsymbol{\upmu }$ denotes the vector of geometric parameters for the U-shaped spring. 

In this study, the weights are assigned as $w_1=0.1$, $w_2=0.5$ and $w_3=0.4$ for vertical bending, lateral bending, and torsion, respectively. The lower weight for vertical bending reflects the significant influence of the cable system on the vertical stiffness of cable-supported truss girders, which reduces sensitivity to discrepancies in the truss girder's vertical bending stiffness. Conversely, lateral bending stiffness strongly governs lateral modal frequencies and aerostatic deformation, justifying a higher weight. Torsional stiffness, critical for flutter instability, is assigned a substantial weight to minimize discrepancies.

As noted in Eq. (\ref{eq:5}), the optimization constraints account for manufacturing precision limitations. Treating geometric parameters as continuous variables often yields dimensions with extended decimal expansions that are impractical for precise machining, leading to rounding errors. Given the small size of U-shaped springs, their stiffness is highly sensitive to discrepancies between designed and manufactured dimensions. To address this, all geometric parameters are quantized to a 0.1 mm resolution, a standard machining tolerance, ensuring that the optimized parameters correspond to feasible manufactured values. Additionally, lower and upper bounds $k_{i,min}$ and $k_{i,max}$  are imposed based on the scaled truss girder's dimensions. Windward dimensions, such as $c, m, L_1, L_2$ (see Fig. \ref{fig:3}), are minimized to reduce aerodynamic interference.

To impose the quantization constraint, geometric variables are rounded to the nearest 0.1 mm at each optimization iteration, i.e., the optimization operates on a discrete grid. This transforms the problem into a non-smooth optimization task where the objective function is non-differentiable at discrete points. Consequently, derivative-free optimization algorithms are employed, including the Nelder-Mead method, Pattern Search method, and Genetic Algorithm, implemented using MATLAB R2023b with the Global Optimization Toolbox \citep{MathWorks2023}. Bound constraints are enforced using the fminsearchbnd method for the Nelder-Mead algorithm \citep{nelder1965simplex,DErrico2006}.

\begin{figure}[H]
    \centering
    \includegraphics[width=1\textwidth]{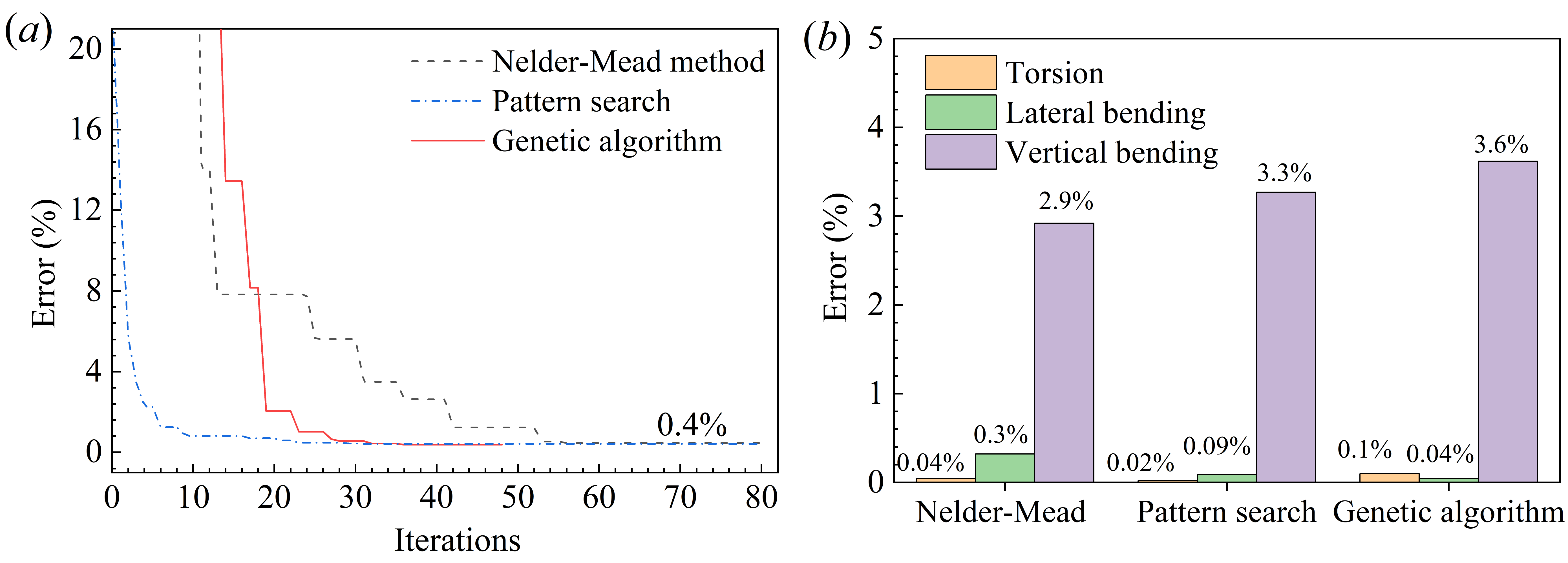}
    \caption{The optimization of design parameters for the truss girder in Fig.~\ref{fig:17}. 
             (\textit{a}) Iterative process, 
             (\textit{b}) Errors of elastic strain energy in different degrees of freedom. 
             The lower and upper bounds are set as 
             $\boldsymbol{\upmu}_{\mathrm{lb}} = [1.0, 1.0, 1.0, 1.0, 8.0, 8.0]^{\mathrm{T}}$ 
             and 
             $\boldsymbol{\upmu}_{\mathrm{ub}} = [5.0, 10.0, 10.0, 10.0, 50.0, 80.0]^{\mathrm{T}}$ 
             for the geometric parameter vector $\boldsymbol{\upmu}$, where the dimensions are in millimeters. 
             The final optimal value is 
             $\boldsymbol{\upmu}_{\mathrm{op}} = [ syst1.0, 3.0, 3.9, 2.1, 48.4, 73.4]^{\mathrm{T}}$.}
    \label{fig:13}
\end{figure}

Fig. \ref{fig:13} illustrates the optimization process for the aeroelastic model of a typical truss girder (see Fig. \ref{fig:17}). The convergence of the objective function during the iterative process is shown in Fig. \ref{fig:13}\textit{a}, demonstrating a rapid reduction in weighted residuals across all three optimization schemes, converging to an optimal value of approximately 0.4\% within seconds. Among the schemes, the Genetic Algorithm proves most robust, offering effective global exploration with a population size of 100 to 1,000, though it exhibits slightly larger residuals for lateral bending (see Fig. \ref{fig:13}\textit{b}). In contrast, the Nelder-Mead and Pattern Search methods are more sensitive to search strategies and may converge to local minima. The performance of the Pattern Search method is enhanced by incorporating a GPS-based search strategy and a larger InitialMeshSize. As depicted in Fig. \ref{fig:13}\textit{a}, the final optimal values of the three schemes are closely aligned, with minor variations in residuals across different degrees of freedom. Given the low computational cost of each scheme, all three are recommended for obtaining an optimal set of design parameters.

\subsection{Numerical validation of modal frequencies} 
\label{sub:numerical_validation_of_modal_frequencies}

\begin{figure}[H]
    \centering
    \includegraphics[width=1\textwidth]{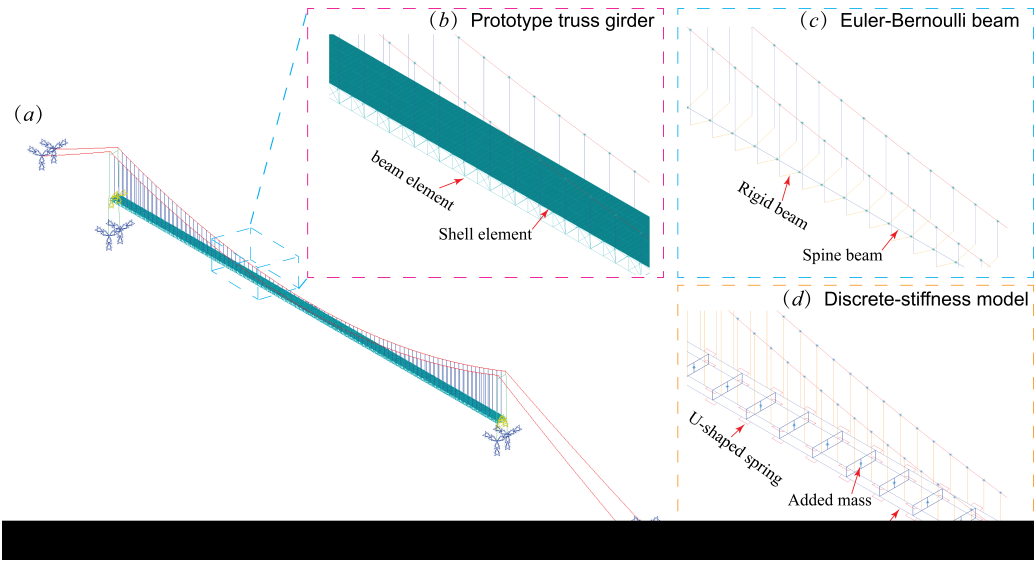}
    \caption{Finite element models of the truss-girder suspension bridge (shown in Fig.~\ref{fig:17}) developed in ANSYS. 
             (\textit{a}) Full bridge model with boundary conditions, 
             (\textit{b}) Multi-scale model of the prototype truss girder,
             (\textit{c}) Spine-beam model representing the equivalent Euler-Bernoulli beam,
             (\textit{d}) Discrete-stiffness model for the scaled aeroelastic main girder.}
    \label{fig:14}
\end{figure}

The optimal design parameters for the U-shaped springs, derived in Section \ref{sub:optimization_algorithm}, are validated through finite element analysis in this section. A finite element model of a representative suspension bridge with a truss girder, as shown in Fig.~\ref{fig:17}, is developed using ANSYS. The main girder is modeled using three distinct approaches, as illustrated in Fig.~\ref{fig:14}:

\begin{enumerate}[label=(\arabic*)]
    \item \textbf{Multi-scale Model:} The prototype truss girder is represented with the bridge deck modeled using shell elements and the truss members using beam elements.
    
    \item \textbf{Spine-beam Model:} The main girder is idealized as an equivalent Euler-Bernoulli beam, employing spine beam elements (Beam4). Cable anchorages are modeled with rigid beam elements, and mass properties are simulated using point mass elements (Mass21).
    
    \item \textbf{Discrete-stiffness Model:} The scaled aeroelastic girder model incorporates U-shaped springs to provide elastic stiffness, rigid blocks to facilitate cable anchorage and constrain the U-shaped springs, and added masses to replicate the mass properties. The U-shaped springs are modeled using both Euler-Bernoulli beam elements (Beam4) and Timoshenko beam elements (Beam188) to evaluate the impact of shear deformation. Rigid blocks are represented by rigid beam elements (Beam4), and the added mass is modeled with point mass elements (Mass21) to simulate the scaled mass and mass moment of inertia for each rigid block.
\end{enumerate}

Given that the simulations of boundary conditions between the main girder and pylons are slightly different for the above three schemes, two scenarios are considered herein: (a) a cantilever main girder only; (b) the whole bridge. The case of cantilever girder is to exclude the influence of detailed simulation of boundary conditions between the main girder and pylons. 

Table \ref{tab:1} presents the modal frequencies of the cantilever aeroelastic model. The relative error $\varepsilon_1$ quantifies the discrepancy arising from approximating the truss girder as an equivalent Euler-Bernoulli beam. The analysis indicates that the lateral frequency exhibits the most significant discrepancies, ranging from 1.5\% to 2.5\%. This suggests that the prototype truss girder cannot be fully idealized as an Euler-Bernoulli beam, with accuracy diminishing as the effective slenderness ratio increases. The effective slenderness ratio is defined as the length-to-width ratio $L/B = 46.6$  for lateral bending modes and the length-to-height ratio $L/D = 177.5$ for vertical bending modes.

The relative error $\varepsilon_2$ reflects discrepancies attributable to the design equations for the U-shaped springs. This error indicates that the discrete stiffness system's accuracy diminishes for higher-order bending modes compared to the first lateral and vertical bending modes, with the largest discrepancy observed in the second lateral mode at -3.2\%.

The relative error $\varepsilon_3$ evaluates the influence of shear deformation in the columns and crossbeams of the U-shaped springs, which were modeled as ideal Euler-Bernoulli beams in the stiffness matrix derivation (Section \ref{sub:stiffness_matrix_of_u_shaped_spring_element}). The results show that shear deformation slightly reduces structural stiffness, but its impact is minimal and can be reasonably neglected.

Similar trends are observed in Table \ref{tab:2}, which reports the calculated modal frequencies for the full bridge aeroelastic model, including the cable system. The relative errors $\varepsilon_1$ and $\varepsilon_2$ are generally smaller compared to the cantilever case, indicating that the cable system mitigates discrepancies. However, the discrepancy for the asymmetric lateral mode increases slightly to -3.3\%, likely due to the pylon constraints at the girder ends, which effectively reduces the effective slenderness ratio compared to the cantilever configuration. Consistent with the cantilever results, the relative error $\varepsilon_3$ due to shear deformation remains negligible.

\begin{table}[H]
    \centering
    \caption{Comparison of modal frequencies for the cantilever aeroelastic model of the truss girder depicted in Fig.~\ref{fig:17}. The symbols are defined as follows: $f_\text{p}$ represents the scaled frequency of the prototype bridge using truss girder model (see Fig.~\ref{fig:14}\textit{b}), where the frequency ratio $\lambda_f = 1 / \sqrt{\lambda_L}$ following Froude number similarity.  $f_\text{EI}$ signifys the scaled frequency of the prototype bridge using the equivalent Euler-Bernoulli beam model (see Fig.~\ref{fig:14}\textit{c}). $f_\text{E}$ is the frequency of the scaled aeroelastic model with U-shaped springs modeled as Euler-Bernoulli beam (Beam4, Fig.~\ref{fig:14}\textit{d}). $f_\text{T}$ is the frequency of the scaled aeroelastic model with U-shaped springs modeled as Timoshenko beams (Beam188, Fig.~\ref{fig:14}\textit{d}). $\varepsilon_1$  represents the relative error due to approximating the truss girder as an equivalent Euler-Bernoulli beam. $\varepsilon_2$ denotes the relative error from the stiffness and design equations of the U-shaped springs. $\varepsilon_3$  reflects the relative error due to the shear deformation in the U-shaped spring components.}
    \footnotesize 
    \setlength{\tabcolsep}{1pt} 
    \begin{tabular}{cccccccc}
        \toprule
        \multirow{2}{*}{Mode shape} & \multicolumn{1}{c}{\text{Truss girder}} & \multicolumn{2}{c}{\text{Euler-Bernoulli beam}} & \multicolumn{4}{c}{\text{Discrete system with U-shaped springs}} \\        
        & $f_\text{p}$ (Hz) & $f_\text{EI}$ (Hz) & $\varepsilon_1 = (f_\text{EI} - f_\text{p}) / f_\text{p}$ & $f_\text{E}$ (Hz) & $\varepsilon_2 = (f_\text{E}-f_\text{p}) / f_\text{p}$ & $f_\text{T}$ (Hz) & $\varepsilon_3 = (f_\text{T} - f_\text{E}) / f_\text{E}$ \\
        \midrule
        1st vertical mode    & 0.0415 & 0.0413 & -0.5\% & 0.0409 & -1.4\% & 0.0397 & -2.93\% \\
        1st lateral mode     & 0.1365 & 0.1387 & 1.6\%  & 0.1367 & 0.1\%  & 0.1357 & -0.75\% \\
        2nd vertical mode    & 0.2594 & 0.2585 & -0.3\% & 0.2539 & -2.1\% & 0.2519 & -0.79\% \\
        3rd vertical mode    & 0.7231 & 0.7241 & 0.1\%  & 0.7035 & -2.7\% & 0.7002 & -0.46\% \\
        2nd lateral mode     & 0.8490 & 0.8701 & 2.5\%  & 0.8219 & -3.2\% & 0.8213 & -0.07\% \\
        1st torsional mode   & 1.3541 & 1.3644 & 0.8\%  & 1.3414 & -0.9\% & 1.3325 & -0.83\% \\
        \bottomrule
    \end{tabular}
    \label{tab:1}
\end{table}

\begin{table}[H]
    \centering
    \caption{Comparison of modal frequencies for the full bridge aeroelastic model of the prototype shown in Fig.~\ref{fig:17}. Symbols are as defined in Table \ref{tab:1}.}
    \footnotesize 
    \setlength{\tabcolsep}{1pt} 
    \begin{tabular}{cccccccc}
        \toprule
        \multirow{2}{*}{Mode shape} & \multicolumn{1}{c}{\text{Truss girder}} & \multicolumn{2}{c}{\text{Euler-Bernoulli beam}} & \multicolumn{4}{c}{\text{Discrete system with U-shaped springs}} \\        
        & $f_\text{p}$ (Hz) & $f_\text{EI}$ (Hz) & $\varepsilon_1 = (f_\text{EI} - f_\text{p}) / f_\text{p}$ & $f_\text{E}$ (Hz) & $\varepsilon_2 = (f_\text{E}-f_\text{p}) / f_\text{p}$ & $f_\text{T}$ (Hz) & $\varepsilon_3 = (f_\text{T} - f_\text{E}) / f_\text{E}$ \\
        \midrule
        S-L-1 & 0.6374 &  0.6359 & -0.2\% & 0.6346 & -0.4\% & 0.6338 & -0.12\% \\
        A-V-1 & 1.2229 & 1.2285 & 0.5\%  & 1.2273 & 0.4\%  & 1.2268 & -0.04\% \\
        A-L-1 & 1.4916 &  1.4877 & -0.3\% & 1.4419 & -3.3\% & 1.4376 & -0.30\% \\
        S-V-1 & 1.7228 &  1.7233 & 0.0\%  & 1.7215 & -0.1\% & 1.7206 & -0.05\% \\
        S-V-2 & 2.2958 &  2.2934 & -0.1\% & 2.2897 & -0.3\% & 2.2889 & -0.03\% \\
        S-T-1 & 3.4763 & 3.4709 & -0.2\% & 3.4567 & -0.6\% & 3.4546 & -0.01\% \\
        A-T-1 & 4.2418 &  4.3513 & 2.6\%  & 4.1750 & -1.6\% & 4.1643 & -0.26\% \\
        \bottomrule
        \multicolumn{8}{l}{\textit{Note:} S: symmetric, A: antisymmetric, V: vertical, L: lateral.} \\
    \end{tabular}
    \label{tab:2}
\end{table}

\subsection{Comparison of aerostatic deformation} 
\label{sub:comparison_of_aerostatic_deformation}
In addition to the modal frequencies analyzed in Section 4.2, the lateral deformation induced by static wind loads is a critical consideration in full-bridge aeroelastic testing. Suspension bridges are designed to limit transverse displacement to ensure comfort and safety for vehicles and pedestrians \citep{MOT2015,AASHTO2008,authority2001wind}. For instance, the Chinese design code for highway suspension bridges requires that the ratio of lateral displacement to span length under wind action be less than 1/150 \citep{MOT2015}. Consequently, achieving accurate simulation of aerostatic deformation, particularly lateral displacement, is essential for truss-girder suspension bridges.

The aerostatic deformation of the truss girder is evaluated using the finite element models described in Fig.~\ref{fig:14}. Static wind loads, corresponding to a reference wind speed of  $U = 44.7$ m/s, are applied to the girder. The wind loads are calculated as:
\begin{equation}
    \label{eq:61}
M\left( {{\alpha _{\rm{0}}}} \right){\rm{ = }}\frac{{\rm{1}}}{{\rm{2}}}\rho {U^{\rm{2}}}{B^{\rm{2}}}{C_{\rm{M}}}\left( {{\alpha _{\rm{0}}}} \right),\;{F_{\rm{H}}}\left( {{\alpha _{\rm{0}}}} \right){\rm{ = }}\frac{{\rm{1}}}{{\rm{2}}}\rho {U^{\rm{2}}}D{C_{\rm{H}}}\left( {{\alpha _{\rm{0}}}} \right),\;{F_{\rm{V}}}\left( {{\alpha _{\rm{0}}}} \right){\rm{ = }}\frac{{\rm{1}}}{{\rm{2}}}\rho {U^{\rm{2}}}B{C_{\rm{V}}}\left( {{\alpha _{\rm{0}}}} \right)
\end{equation}
where $\alpha_0$ represents the initial wind angle of attack. $\rho$ denotes the air density, and $U$ is the mean wind speed. $B,D$ represent the width and height of the girder, respectively (see Fig.~\ref{fig:17}). $F_\text{H},F_\text{V},M$ are, respectively, the aerostatic horizontal force, vertical force, and torsional moment per unit length in the body-axis coordinate system. The non-dimensional force coefficients $C_\text{H},C_\text{V},C_\text{M}$ are determined in wind-tunnel section model tests.

\begin{figure}[H]
    \centering
    \includegraphics[width=1\textwidth]{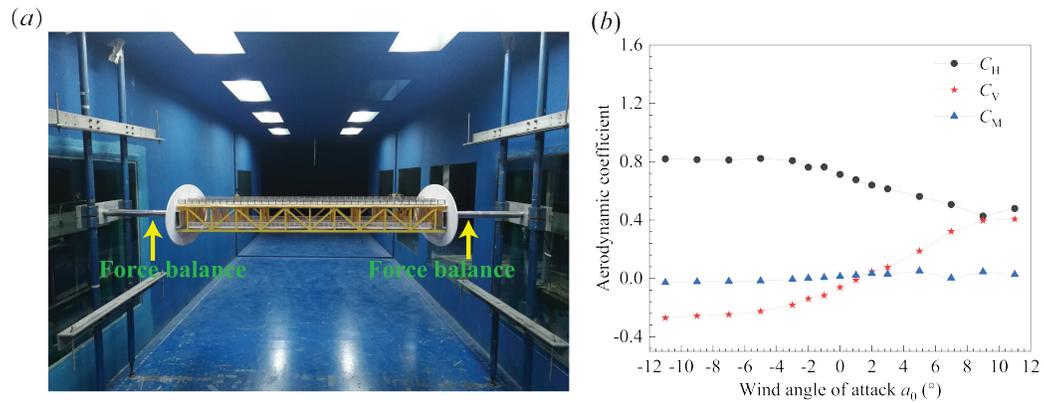}
    \caption{Aerostatic coefficients of a typical truss deck (Fig.~\ref{fig:17}). 
             (\textit{a}) Experimental setup of wind-tunnel section model tests, 
             (\textit{b}) Measured aerodynamic coefficients versus wind angle of attack.}
    \label{fig:15}
\end{figure}

The measurement of aerostatic force coefficients is illustrated in Fig.~\ref{fig:15}. A section model, replicating the geometric details of the prototype truss girder, was constructed and tested in the CA-1 wind tunnel test section. Aerostatic forces were measured using five-axis strain gauge balances (Fig.~\ref{fig:15}\textit{a}) and projected onto the body-axis coordinate system to derive the force coefficients. As shown in Fig.~\ref{fig:15}\textit{b}, the aerostatic force coefficients vary with the wind angle of attack.

\begin{figure}[H]
    \centering
    \includegraphics[width=1\textwidth]{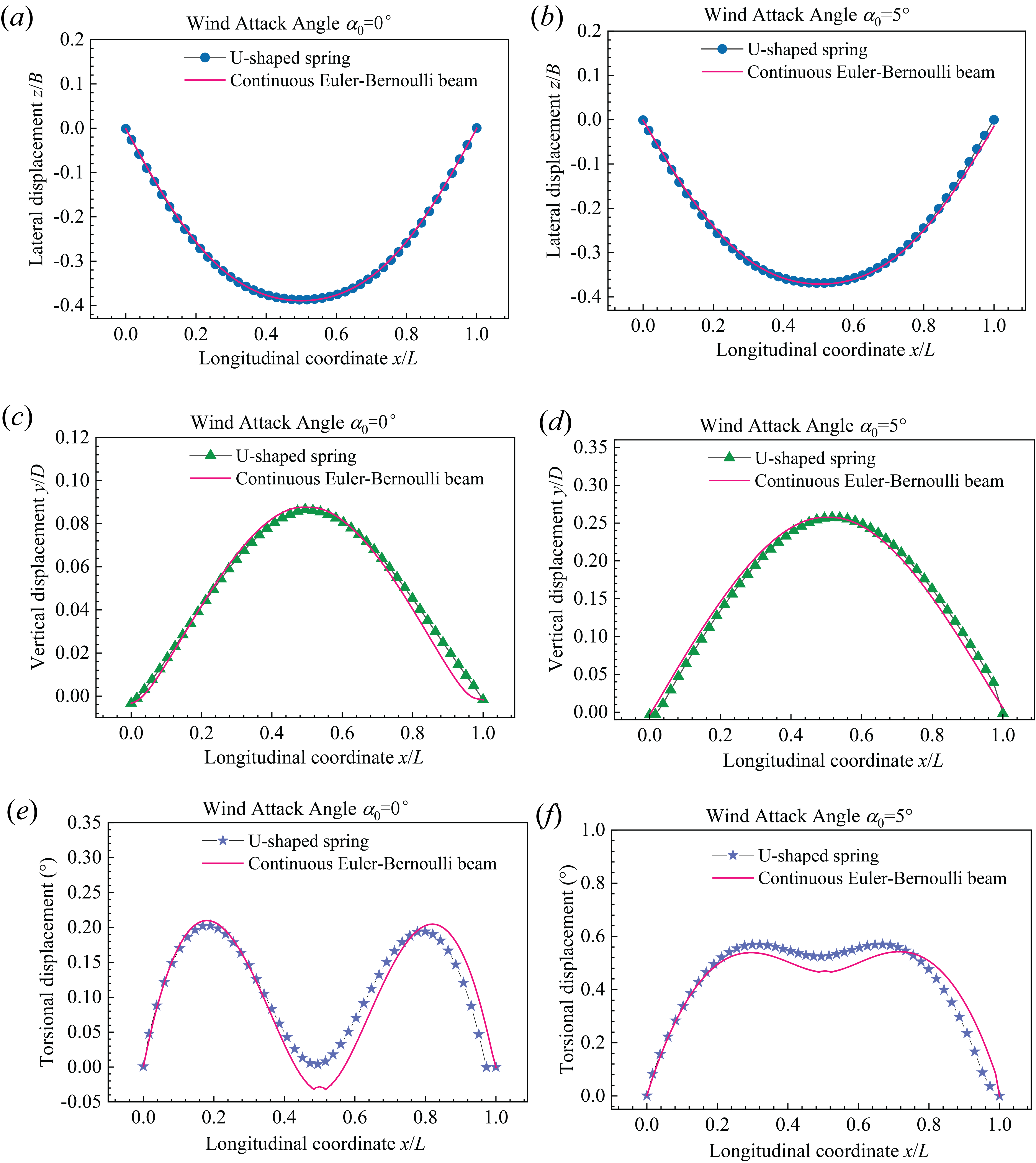}
    \caption{Static deformation of truss girders for a long-span suspension bridge.}
    \label{fig:16}
\end{figure}

Nonlinear static analysis in ANSYS, incorporating large deformation and stress-stiffening effects, is conducted to compute the aerostatic deformation. Two wind angles of attack, $0^\circ$ and $5^\circ$, are considered. The results, presented in Fig.~\ref{fig:16}, compare the deformation of the main girder modeled using the equivalent Euler-Bernoulli beam (Fig.~\ref{fig:14}\textit{c}) and the discrete-stiffness system with U-shaped springs (Fig.~\ref{fig:14}\textit{d}). The discrete-stiffness model demonstrates excellent agreement with the Euler-Bernoulli beam model.

These findings confirm that the geometric parameters of the U-shaped springs, optimized as described in Section \ref{sub:optimization_algorithm}, accurately simulate the stiffness of the truss girder, thereby validating their suitability for aeroelastic modeling.


\section{Experimental case study} 
\label{sec:experimental_case_study}

This section presents the design, fabrication, and experimental validation of an aeroelastic model for a typical suspension bridge, following the design procedure outlined in Section \ref{sub:discrete_stiffness_system_and_design_procedure_of_u_shaped_springs}. The engineering context and design parameters are introduced in Section \ref{sub:a_long_span_suspension_bridge_with_a_truss_girder}. The fabrication process for the aeroelastic model is detailed in Section \ref{sub:fabrication_of_the_aeroelastic_model}. The accuracy of the stiffness matrices for the U-shaped spring and spring-lever elements, derived in Sections 3.1 and 3.3, is evaluated in Section \ref{sub:stiffness_matrix_validation_for_u_shaped_spring_and_spring_lever_elements}. Finally, the modal frequencies of the aeroelastic model are experimentally measured in Section \ref{sub:measurement_of_modal_frequencies} to quantify errors arising from fabrication and installation.

\subsection{A long-span suspension bridge with a truss girder} 
\label{sub:a_long_span_suspension_bridge_with_a_truss_girder}

\begin{figure}[H]
    \centering
    \begin{subfigure}[b]{1\textwidth}
        \centering
        \includegraphics[width=1\textwidth]{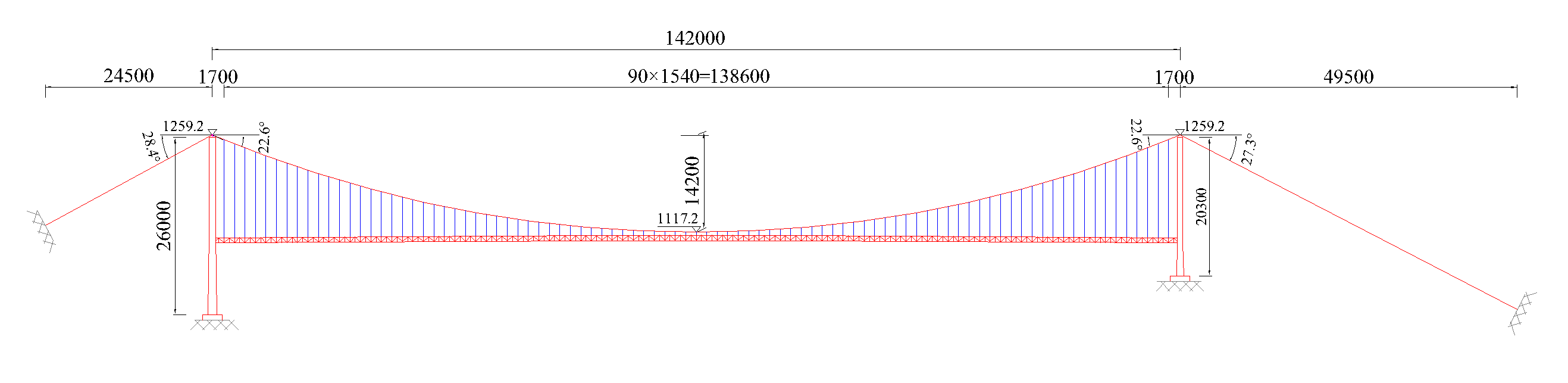}
        \caption{Elevation layout (unit: cm)}
    \end{subfigure}
    \hfill
    \begin{subfigure}[b]{0.8\textwidth}
        \centering
        \includegraphics[width=1\textwidth]{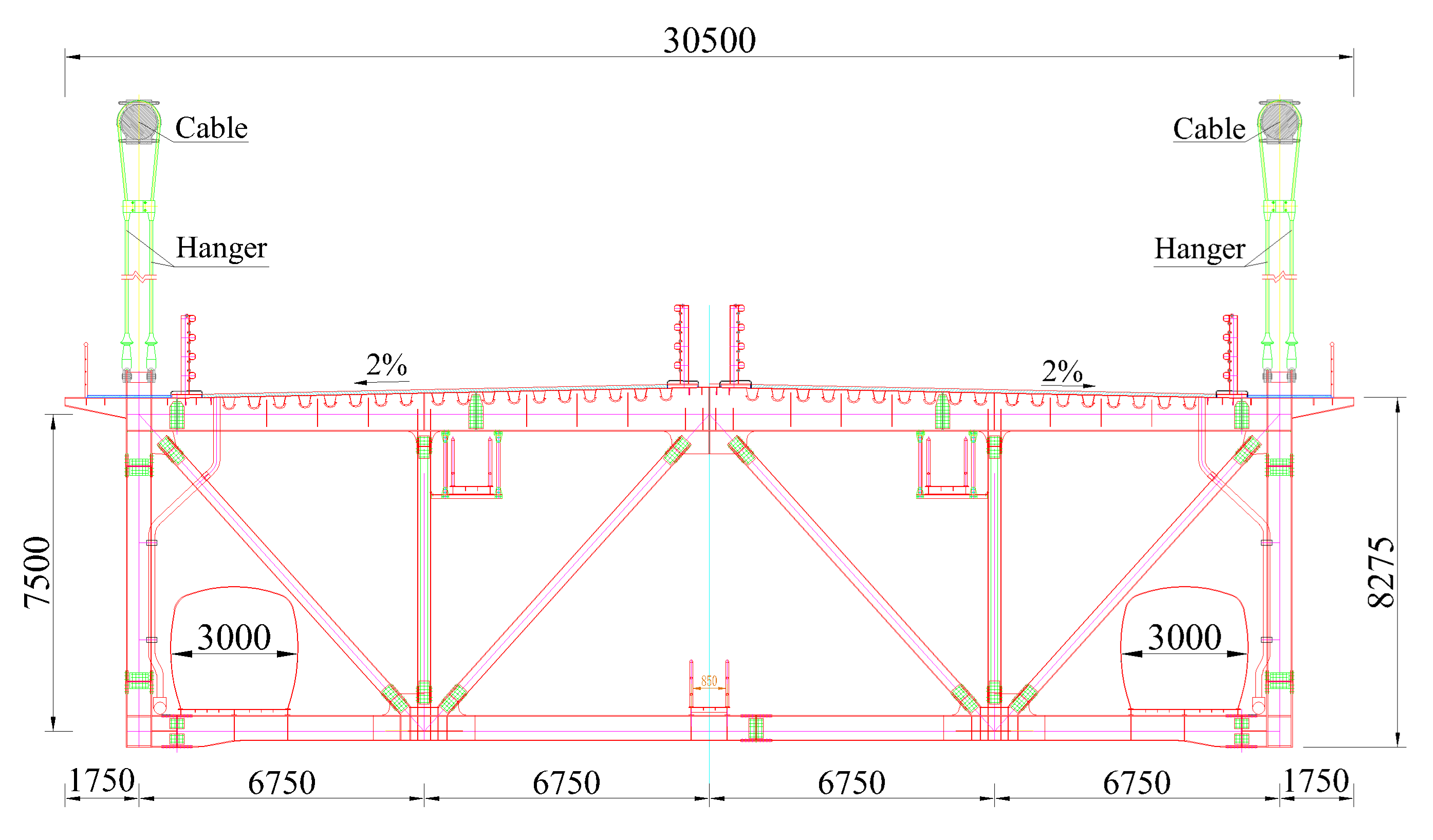}
        \caption{Cross section of truss girder (unit: mm)}
    \end{subfigure}
       \caption{The structural dimensions of a single-span suspension bridge with a truss girder.}
       \label{fig:17}
\end{figure}

The Huajiang Canyon Bridge, featuring a main span of 1,420 m and a truss girder configuration (see Fig.~\ref{fig:17}), serves as the engineering background for this study. Its primary structural parameters are summarized in Table \ref{tab:3}. A geometric length scale of  $\lambda_L=1:173$ is selected to accommodate the dimensions of the CA-3 wind tunnel test section at Chang'an University and to ensure compatibility with the diameter of steel wires used to simulate the main cable.

The design of the full-bridge aeroelastic model adheres to similarity principles for geometry, mass, and elastic stiffness of the main structural components. Based on the Froude number similarity discussed in Section \ref{sub:scaling_principles}, the velocity and frequency scaling factors are derived from the length scale as $\lambda_U=1:\sqrt{173}$ and $\lambda_f=\sqrt{173}:1$, respectively. Using these scaling relationships, the mass and stiffness parameters for all structural components are determined, with those for the main girder listed in Table \ref{tab:3}.

The scaled geometry and elastic stiffness properties of the truss girder, as provided in Table \ref{tab:3}, are used to compute the geometric parameters of the U-shaped springs through the design equations and optimization procedure described in Sections \ref{sub:design_equations_by_elastic_strain_energy_equivalence} and \ref{sub:optimization_algorithm}. In addition to the geometric constraints outlined in Section \ref{sub:optimization_algorithm}, an additional fabrication constraint is imposed to facilitate fabrication: the out-of-plane cross-sectional dimension of the crossbeam is set equal to that of the column, i.e., $n=d$. The U-shaped springs are fabricated from steel with Young's modulus $E=2.06 \times 10^{11}$ N/m² and shear modulus $G=7.92 \times 10^{10}$ N/m². The resulting optimal geometric parameter vector is ${\boldsymbol{\rm{\upmu}}_{{\rm{op}}}} = {[\begin{array}{*{20}{c}}{1.0}&{1.0}&{4.0}&{1.0}&{15.0}&{50.0}\end{array}]^{\rm{T}}}$, with dimensions expressed in millimeters.

\begin{table}[H]
    \centering
    \caption{Design parameters of the full bridge aeroelastic model. The mass moment of inertia $I_\text{p}$, elastic vertical bending stiffness $EI_z$, lateral bending stiffness $EI_y$, and torsional stiffness $GJ_\text{d}$ correspond to the equivalent Euler-Bernoulli beam, derived from the truss girder using the elastic strain energy equivalence method described in Section \ref{sub:equivalent_euler_bernoulli_beam_by_elastic_strain_energy_equivalence}.}
    \footnotesize 
    \setlength{\tabcolsep}{1pt} 
    \begin{tabular}{ccccc}
        \toprule
        Components & Parameter & Prototype bridge & Scaling ratio  & Aeroelastic model \\
        \midrule
        Total length of truss girder & $L$ (m) & 1420 & $\lambda_L = 1/173$ & 8.208 \\ 
        Width of truss girder & $B$ (m) & 30.5 & $\lambda_L = 1/173$ & 0.176 \\ 
        Height of truss girder & $D$ (m) & 8.0 & $\lambda_L = 1/173$ & 0.046 \\ 
        Mass per unit length of truss girder & $m_p$ (kg/m) & 32913.6 & $\lambda_L^2 = 1/173^2$ & 1.099 \\ 
        Mass moment inertia per unit length of the girder & $I_p$ (kg$\cdot$m$^2$/m) & $1.619 \times 10^6$ & $\lambda_L^4 = 1/173^4$ & $1.807 \times 10^{-3}$ \\ 
        Lateral bending stiffness of the girder & $EI_y$ (N$\cdot$m$^2$) & $2.069 \times 10^{13}$ & $\lambda_L^5 = 1/173^5$ & 133.515 \\
        Vertical bending stiffness of the girder & $EI_z$ (N$\cdot$m$^2$) & $1.818 \times 10^{12}$ & $\lambda_L^5 = 1/173^5$ & 11.732 \\ 
        Torsional stiffness of the girder & $GJ_d$ (N$\cdot$m$^2$) & $5.474 \times 10^{11}$ & $\lambda_L^5 = 1/173^5$ & 3.532 \\ 
        \bottomrule
        \end{tabular}
    \label{tab:3}
\end{table}

\subsection{Fabrication of the aeroelastic model} 
\label{sub:fabrication_of_the_aeroelastic_model}

\begin{figure}[H]
    \centering
    \includegraphics[width=1\textwidth]{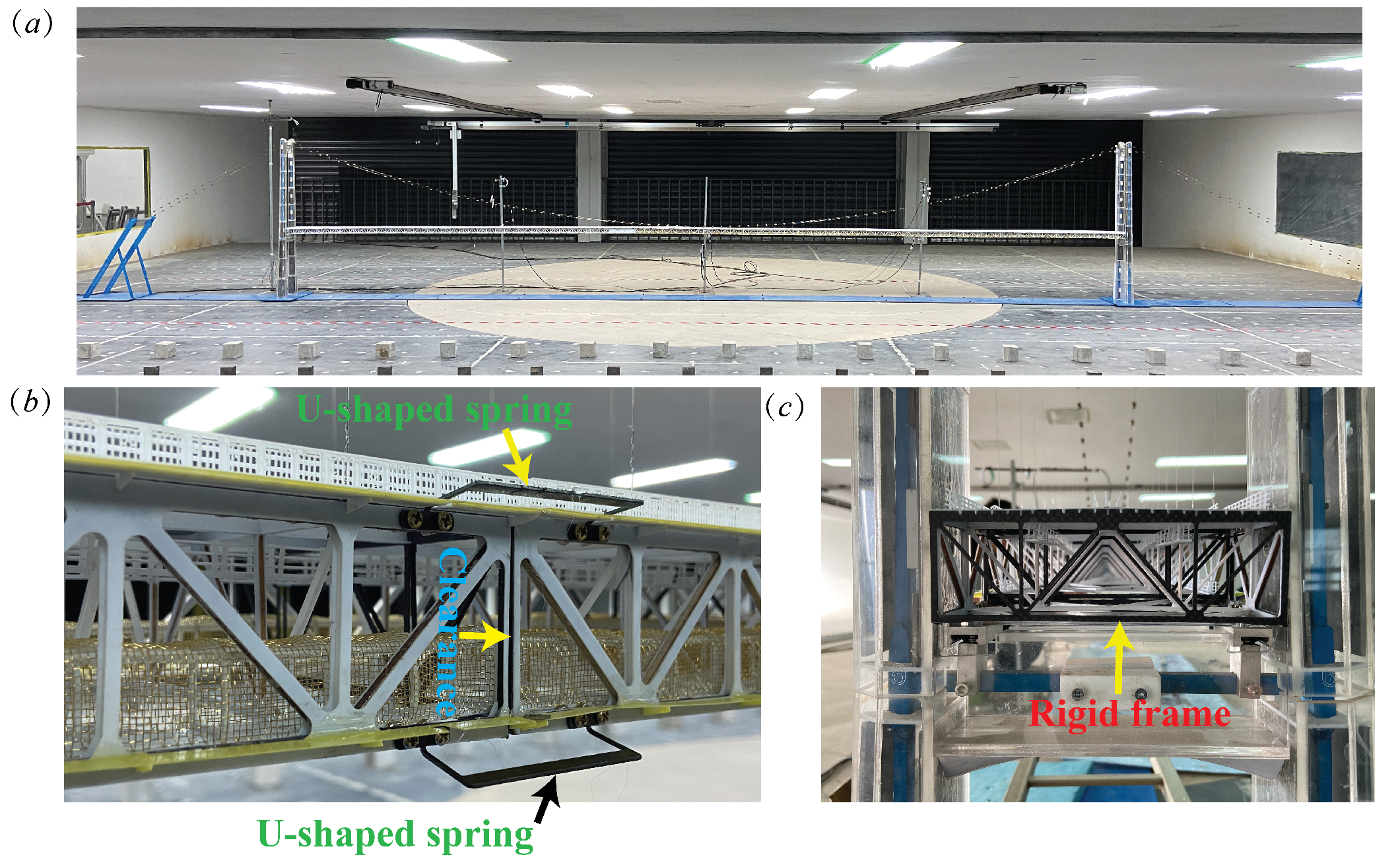}
    \caption{Experimental setup of the full bridge aeroelastic model in the CA-3 wind tunnel. (\textit{a}) General elevation view, (\textit{b}) Rigid segments connected end-to-end via U-shaped springs, (\textit{c}) Side view of rigid frames.}
    \label{fig:18}
\end{figure}

The fabricated aeroelastic model is depicted in Fig. \ref{fig:18}. During fabrication, the geometric similarity of the structural components, particularly the truss girder, is meticulously maintained to ensure accurate representation of the prototype. Fig. \ref{fig:18}\textit{a} illustrates the overall layout and installation of the model within the wind tunnel. Detailed views of the girder segments interconnected by U-shaped springs are shown in Fig. \ref{fig:18}\textit{b}. To enhance the rigidity of each girder segment, a metal rigid frame is incorporated to reinforce the truss system, as presented in Fig. \ref{fig:18}\textit{c}.

As shown in Fig. \ref{fig:18}\textit{b}, clearances of 1 mm are positioned at the midpoints between adjacent hanger intervals, with each girder segment spanning two hanger intervals. Consequently, the main girder is divided into 46 segments, each with a standard length of $l=0.178$m. Given that the number of segments $n=46$ is relatively low, and referencing the findings discussed in relation to Fig. \ref{fig:12}, the implicit forms of the design equations (Eqs. (\ref{eq:48}) and (\ref{eq:55})) are utilized to determine the geometric parameters of the U-shaped springs.

\subsection{Stiffness matrix validation for U-shaped spring and spring-lever elements} 
\label{sub:stiffness_matrix_validation_for_u_shaped_spring_and_spring_lever_elements}

Using the design parameters established for the aeroelastic model, the accuracy of the stiffness matrices ${\left[ {{{\bf{K}}_{\rm{u}}}} \right]^{\rm{e}}}$ dervied for the U-shaped spring in Section \ref{sub:stiffness_matrix_of_u_shaped_spring_element}, and  ${\left[ {{{\bf{K}}}} \right]^{\rm{e}}}$ of the spring-lever element in Section \ref{sub:stiffness_matrix_of_a_cantilever_spring_lever_element} is evaluated.

The theoretical stiffness matrix ${\left[ {{{\bf{K}}_{\rm{u}}}} \right]^{\rm{e}}}$ for the U-shaped spring is computed by substituting the optimized geometric parameters $\boldsymbol{\upmu}_{\mathrm{op}}$ from Section \ref{sub:a_long_span_suspension_bridge_with_a_truss_girder} into Eq. (\ref{eq:15}), yielding: 
\[\left[ {{{\bf{K}}_{\rm{u}}}} \right]_{{\rm{theory}}}^{\rm{e}} = \left[ {\begin{array}{*{20}{c}}
{28323.80}&0&0&0&0&{ - 217.81}\\
0&{906.53}&0&0&0&{ - 22.66}\\
0&0&{539.79}&{4.99}&{13.49}&0\\
0&0&{4.99}&{0.48}&{0.13}&0\\
0&0&{13.49}&{0.12}&{0.63}&0\\
{ - 217.82}&{ - 22.66}&0&0&0&{2.79}
\end{array}} \right]\]

\begin{figure}[H]
    \centering
    \includegraphics[width=1\textwidth]{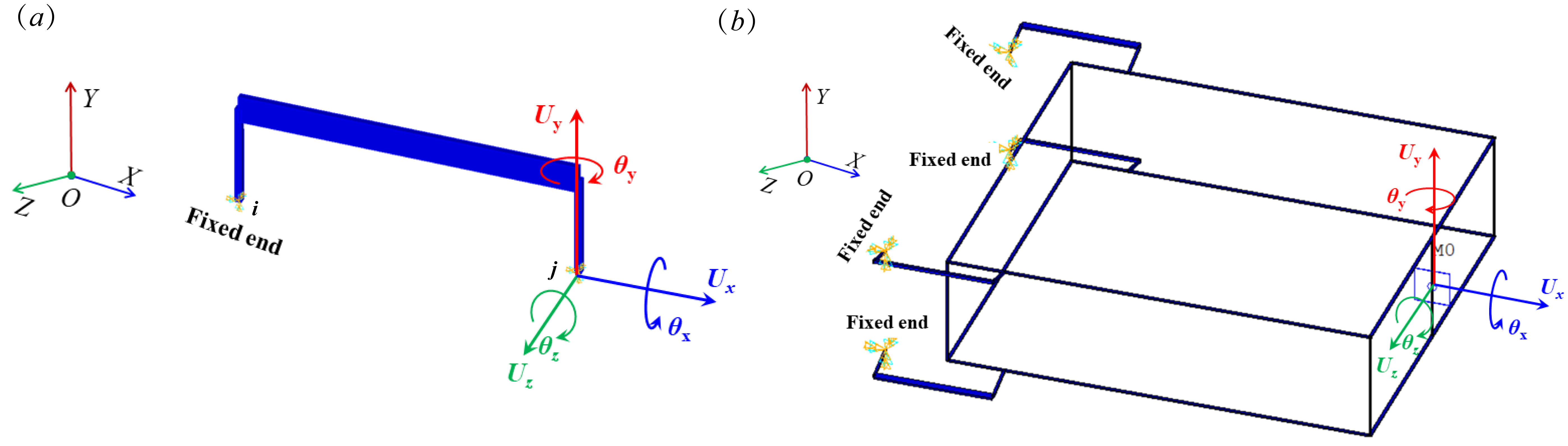}
    \caption{Finite element modeling for stiffness matrix calculation using the unit displacement method. (\textit{a}) U-shaped spring element, (\textit{b}) Spring-lever element.}
    \label{fig:19}
\end{figure}

This stiffness matrix ${\left[ {{{\bf{K}}_{\rm{u}}}} \right]^{\rm{e}}}$ for the U-shaped spring is further calculated via the finite element model, as depicted in Fig. \ref{fig:19}\textit{a}, using unit displacement method. That involves fixing the lower ends of the spring's columns, and applying a unit displacement at node $j$. The resulting reaction forces at node $j$, obtained through static analysis, correspond to one column of the stiffness matrix ${\left[ {{{\bf{K}}_{\rm{u}}}} \right]^{\rm{e}}}$. The calculated stiffness matrix using the finite element method is expressed as 
\[\left[ {{{\bf{K}}_{\rm{u}}}} \right]_{{\rm{FEM}}}^{\rm{e}} = \left[ {\begin{array}{*{20}{r}}
{28310.47}&0&0&0&0&{ - 217.71}\\
0&{907.4}&0&0&0&{ - 22.68}\\
0&0&{528.70}&{4.93}&{13.21}&0\\
0&0&{4.93}&{0.47}&{0.12}&0\\
0&0&{13.21}&{0.12}&{0.623}&0\\
{ - 217.71}&{ - 22.68}&0&0&0&{2.8}
\end{array}} \right]\]

Analogously, the stiffness matrix ${\left[ {{{\bf{K}}}} \right]^{\rm{e}}}$ of a spring-lever element can obtained by substituting the $\boldsymbol{\upmu}_{\mathrm{op}}$ into Eq.(\ref{eq:43}) as
\[\left[ {\bf{K}} \right]_{{\rm{theory}}}^{\rm{e}} = \left[ {\begin{array}{*{20}{r}}
{113295.000}&0&0&0&0&0\\
0&{3626.130}&0&0&0&{ - 736.104}\\
0&0&{2159.170}&0&{438.312}&0\\
0&0&0&{26.072}&0&0\\
0&0&{438.312}&0&{779.455}&0\\
0&{ - 736.104}&0&0&0&{259.000}
\end{array}} \right]\]

The stiffness matrix can also be numerically calculated using using unit displacement method via finite element model of a spring-elver element (see Fig. \ref{fig:19}\textit{b}), which is 
\[\left[ {\bf{K}} \right]_{{\rm{FEM}}}^{\rm{e}} = \left[ {\begin{array}{*{20}{r}}
{112355.469}&0&0&0&0&0\\
0&{3625.000}&0&0&0&{ - 734.524}\\
0&0&{2113.281}&0&{425.924}&0\\
0&0&0&{25.391}&0&0\\
0&0&{425.924}&0&{767.578}&0\\
0&{ - 734.524}&0&0&0&{257.812}
\end{array}} \right]\]

\begin{figure}[H]
    \centering
    \includegraphics[width=1\textwidth]{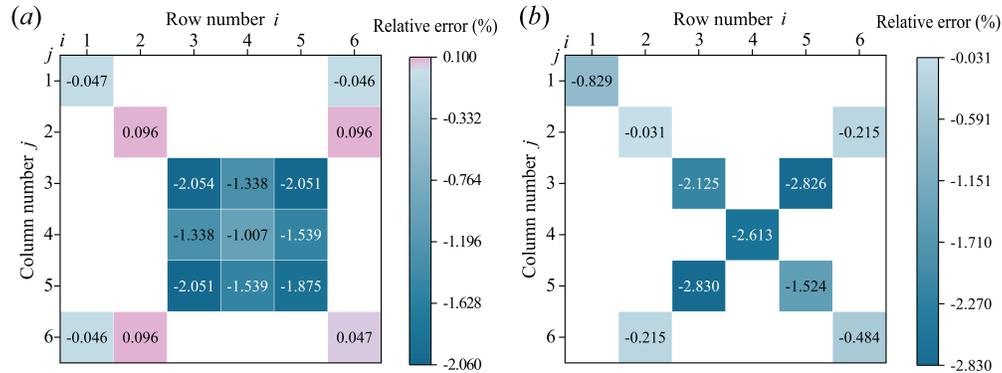}
    \caption{Relative errors between theoretical and finite element-derived stiffness matrices. (\textit{a}) Stiffness matrix $\left[ {{{\bf{K}}_{\rm{u}}}} \right]^{\rm{e}}$ of the U-shaped spring, (\textit{b}) Stiffness matrix of the spring-lever element $\left[ {{{\bf{K}}}} \right]^{\rm{e}}$.}
    \label{fig:20}
\end{figure}

A comparison between the theoretical stiffness matrix $\left[ {{{\bf{K}}_{\rm{u}}}} \right]_{{\rm{theory}}}^{\rm{e}}$ and the numerical stiffness matrix $\left[ {{{\bf{K}}_{\rm{u}}}} \right]_{{\rm{FEM}}}^{\rm{e}}$ reveals close agreement. Similarly, the stiffness matrix $\left[ {\bf{K}} \right]_{{\rm{theory}}}^{\rm{e}}$ for the spring-lever element is validated against its finite element counterpart $\left[ {\bf{K}} \right]_{{\rm{FEM}}}^{\rm{e}}$. The relative errors for each matrix term are presented in Fig. \ref{fig:20}. The closed-form solutions for the stiffness matrices demonstrate satisfactory accuracy, with relative errors below 2.1\% for the U-shaped spring and 2.83\% for the spring-lever element. 

The observed discrepancies are likely attributable to the approximate formula for the torsional constant used in Eq. (\ref{eq:20}), which typically introduces an error of up to 4\% \citep{young2002roark}. These errors could be further reduced by adopting a higher-precision approximation for the torsional constant in place of Equation (\ref{eq:20}).


\subsection{Measurement of modal frequencies} 
\label{sub:measurement_of_modal_frequencies}

Following the installation of the aeroelastic model in the CA-3 wind tunnel, free vibration tests were conducted by manually exciting the model. Modal frequencies were extracted from the free-decay responses, recorded using accelerometers and laser displacement sensors positioned at the quarter-span and mid-span sections of the model.

The measured modal frequencies are presented in Table \ref{tab:4}. The relative error $\varepsilon_1$, which quantifies the discrepancy arising from the design parameters of the U-shaped springs in approximating the elastic stiffness of the prototype truss girder, is minimal across all tested modes. This confirms the high accuracy of the design parameters, as determined through the optimization algorithm described in Section \ref{sub:optimization_algorithm}. In contrast, the relative error $\varepsilon_2$, attributed to fabrication and installation processes of the aeroelastic model, is more significant and constitutes the primary source of error. Nevertheless, this discrepancy remains below 5\%, which is generally acceptable for engineering applications. These findings underscore the importance of precision in manufacturing and installation to minimize errors. The observed discrepancies in modal frequencies may affect the vibration responses of the aeroelastic model. To address this, experimental results should be adjusted to account for these errors, for instance, by applying the correction method proposed by Lan et al. \citep{lan2024new}.

\begin{table}[H]
    \centering
    \caption{Comparison of modal frequencies of the full bridge aeroelastic model. $f_\text{p}$ and $f_\text{E}$ are as defined in Table \ref{tab:1}. $f_\text{m}$ represents the measured frequency}
    \footnotesize 
    \setlength{\tabcolsep}{1pt} 
    \begin{tabular}{cccccccc}
        \toprule
        \multirow{2}{*}{Mode shape} & \multicolumn{1}{c}{\text{Scale prototype bridge}} & \multicolumn{2}{c}{\text{Discrete system with U-shaped springs}} & \multicolumn{2}{c}{\text{Experimental measurement}} \\        
        & $f_\text{p}$ (Hz) & $f_\text{E}$ (Hz) & $\varepsilon_1 = (f_\text{E} - f_\text{p}) / f_\text{p}$ & $f_\text{m}$ (Hz) & $\varepsilon_2 = (f_\text{m}-f_\text{E}) / f_\text{E}$  \\
        \midrule
        S-L-1 & 0.6374 & 0.6443 & 1.1\%  &  0.6510 & 1.0\% \\
        A-V-1 &1.2229 & 1.2337 & 0.9\%   & 1.2719 & 3.1\% \\
        A-L-1 &1.4916 & 1.5061 & 1.0\% &   1.5625 & 3.7\%  \\
        S-V-1 &1.7228 & 1.7317 & 0.5\%  &  1.7578 & 1.5\% \\
        A-V-2 & 2.2959 & 2.3005 & 0.2\%  &  2.3438 & 1.9\% \\
        S-T-1 & 3.4764 & 3.4635 & -0.4\% &  3.3200 & -4.1\% \\
        A-T-1 & 4.2418 & 4.3357 & 2.2\%  &  4.3945 & 1.4\% \\
        \bottomrule
        \multicolumn{6}{l}{\textit{Note:} S: symmetric, A: antisymmetric, V: vertical, L: lateral.} \\
    \end{tabular}
    \label{tab:4}
\end{table}

\section{Conclusions} 
\label{sec:conclusions}

Aeroelastic model testing of truss-girder suspension bridges is a fundamental tool for evaluating aerodynamic performance and mitigating potential wind-induced vibrations. The aeroelastic modeling of truss girders typically employs a discrete-stiffness system with U-shaped springs to replicate the scaled elastic properties of the prototype truss girder. Despite its widespread use, the lack of an efficient design method for U-shaped springs has posed significant challenges in practical application. Conventional design technique relies on a computationally intensive trial-and-error process to evaluate all possible geometric combinations.

This work proposes a novel design procedure to streamline the design of U-shaped springs. Initially, the supporting condition of the prototype truss girder is simplified to a cantilever configuration. Using this cantilever truss girder, the elastic stiffness properties—such as the bending moment of inertia and torsional constant—are determined based on the principle of elastic-strain energy equivalence. A closed-form stiffness matrix for U-shaped springs is then derived using the classical unit displacement method. This matrix is incorporated into a set of design equations that correlate the geometric parameters of the U-shaped springs with the scaled stiffness coefficients of the truss girder.

Consequently, the design for U-shaped springs is transformed into a constrained optimization problem. Accounting for limitations in manufacturing precision, it is further formulated as a non-smooth optimization problem, for which derivative-free algorithms---such as the Nelder-Mead method, Pattern Search method, and Genetic Algorithm---are recommended. The accuracy and feasibility of the derived geometric parameters are validated through numerical and experimental case studies of a typical suspension bridge with a truss girder.

In summary, the proposed design method for U-shaped springs is straightforward to implement and achieves rapid convergence to optimal design parameters within seconds. When applied to the aeroelastic modeling of truss-type suspension bridges, discrepancies in modal frequencies primarily arise from model fabrication and assembly processes instead of the design parameters. Notably, this method is primarily applicable to suspension bridges, and its suitability for truss-type cable-stayed bridges warrants further investigation due to potential complexities introduced by simulating axial stiffness in the design equations. The proposed approach can also be extended to the aeroelastic modeling of other truss-type structures, such as transmission towers and arch bridges.


\section*{Acknowledgment}

The research was supported by National Natural Science Foundation of China, 52278478 and fundamental Research Funds for the Central Universities, CHD (No. 300102214914).

\section*{CRediT authorship contribution statement}

\textbf{Guangzhong Gao}: Writing-Original Draft, Writing-Editing, Conceptualization, Software, Methodology, Formal analysis, Funding Acquisition. \textbf{Wenkai Du}: Writing-Original Draft, Software, Writing-Editing, Writing-Review. \textbf{Yanbo Sun}: Software, Validation, Methodology, Formal analysis. \textbf{Yonghui Xie}: Methodology, Validation, Writing-Review. \textbf{Jiawu Li}: Supervision, Writing-Review. \textbf{Ledong Zhu}: Supervision, Writing-Review.

\section*{Declaration of Competing Interest}

The authors declare that they have no competing financial interests of personal relationships that could have appeared to influence the work reported in this paper.

\section*{Data Availability}

Data will be made available on request.

\bibliographystyle{elsarticle-num-names} 
\bibliography{cas-refs}






\section*{Appendix A: Derived stiffness matrix of a U-shaped spring element}
\label{Appendix_A}

The stiffness matrix of a U-shaped spring element are expressed as follows
\begin{align}
{\bf{\bar F}}_j^{\rm{e}} &= {\left[ {{{\bf{K}}_{\rm{u}}}} \right]^{\rm{e}}}{\bf{\bar U}}_j^{\rm{e}} \tag{12} \\
{\left[ {{{\bf{K}}_\text{u}}} \right]^{\rm{e}}} &= \left[ {\begin{array}{*{20}{c}}
{K_{11}^{\rm{e}}}&0&0&0&0&{K_{16}^{\rm{e}}}\\
0&{K_{22}^{\rm{e}}}&0&0&0&{K_{26}^{\rm{e}}}\\
0&0&{K_{33}^{\rm{e}}}&{K_{34}^{\rm{e}}}&{K_{35}^{\rm{e}}}&0\\
0&0&{K_{43}^{\rm{e}}}&{K_{44}^{\rm{e}}}&{K_{45}^{\rm{e}}}&0\\
0&0&{K_{53}^{\rm{e}}}&{K_{54}^{\rm{e}}}&{K_{55}^{\rm{e}}}&0\\
{K_{61}^{\rm{e}}}&{K_{62}^{\rm{e}}}&0&0&0&{K_{66}^{\rm{e}}}
\end{array}} \right] \tag{15}
\end{align}
where each component is derived and expressed as follows 
\begin{align}
K_{_{11}}^{\rm{e}} &= \frac{{3{i_1}\left( {{i_1} + 2{i_2}} \right)}}{{\left( {2{i_1} + {i_2}} \right)L_1^2}} \tag{A--1} \\
K_{_{16}}^{\rm{e}} &= K_{_{61}}^{\rm{e}} = \frac{{ - 3{i_1}\left( {{i_1} + {i_2}} \right)}}{{\left( {2{i_1} + {i_2}} \right){L_1}}} \tag{A--2}\\
K_{22}^{\rm{e}} &= \frac{{12{i_1}{i_2}}}{{({i_1} + 6{i_2}){L_2}^2}} \tag{A--3}\\
K_{26}^{\rm{e}} &= K_{62}^{\rm{e}} =  - \frac{{6{i_1}{i_2}}}{{\left( {{i_1} + 6{i_2}} \right){\rm{ }}{L_2}}} \tag{A--4}\\
K_{33}^{\rm{e}} &= \frac{{12{j_1}{i_3}{i_4}(2{j_2} + {i_3})}}{{6{i_3}{i_4}{L_2}^2(2{j_2} + {i_3}) + {j_1}\left( {4{j_2}{i_4}{L_1}^2 + 8{i_3}{i_4}{L_1}^2 + 2{j_2}{i_3}{L_2}^2 + {i_3}^2{L_2}^2} \right)}} \tag{A--5}\\
K_{34}^{\rm{e}} &= K_{43}^{\rm{e}} = \frac{{12{j_1}{i_3}{i_4}({j_2} + {i_3}){L_1}}}{{6{i_3}{i_4}{L_2}^2(2{j_2} + {i_3}) + {j_1}\left( {4{j_2}{i_4}{L_1}^2 + 8{i_3}{i_4}{L_1}^2 + 2{j_2}{i_3}{L_2}^2 + {i_3}^2{L_2}^2} \right)}} \tag{A--6}\\
K_{35}^{\rm{e}} &= K_{53}^{\rm{e}} = \frac{{6{j_1}{i_3}{i_4}(2{j_2} + {i_3}){L_2}}}{{6{i_3}{i_4}{L_2}^2(2{j_2} + {i_3}) + {j_1}\left( {4{j_2}{i_4}{L_1}^2 + 8{i_3}{i_4}{L_1}^2 + 2{j_2}{i_3}{L_2}^2 + {i_3}^2{L_2}^2} \right)}} \tag{A--7}\\
K_{44}^{\rm{e}} &= \frac{{{i_3}\left[ {6{j_2}{i_3}{i_4}{L_2}^2 + {j_1}\left( {8{j_2}{i_4}{L_1}^2 + 12{i_3}{i_4}{L_1}^2 + {j_2}{i_3}{L_2}^2} \right)} \right]}}{{6{i_3}{i_4}{L_2}^2(2{j_2} + {i_3}) + {j_1}\left( {4{j_2}{i_4}{L_1}^2 + 8{i_3}{i_4}{L_1}^2 + 2{j_2}{i_3}{L_2}^2 + {i_3}^2{L_2}^2} \right)}} \tag{A--8}\\
K_{55}^{\rm{e}} &= \frac{{4{j_1}{i_4}\left[ {3{i_3}{i_4}{L_2}^2(2{j_2} + {i_3}) + {j_1}\left( {{j_2}{i_4}{L_1}^2 + 2{i_3}{i_4}{L_1}^2 + 2{j_2}{i_3}{L_2}^2 + {i_3}^2{L_2}^2} \right)} \right]}}{{({j_1} + 2{i_4})\left[ {6{i_3}{i_4}{L_2}^2(2{j_2} + {i_3}) + {j_1}\left( {4{j_2}{i_4}{L_1}^2 + 8{i_3}{i_4}{L_1}^2 + 2{j_2}{i_3}{L_2}^2 + {i_3}^2{L_2}^2} \right)} \right]}} \tag{A--9}\\
K_{45}^{\rm{e}} &= K_{54}^{\rm{e}} = \frac{{6{j_1}{i_3}{i_4}{L_1}{L_2}\left( {{j_2} + {i_3}} \right)}}{{6{i_3}{i_4}{L_2}^2(2{j_2} + {i_3}) + {j_1}\left( {4{j_2}{i_4}{L_1}^2 + 8{i_3}{i_4}{L_1}^2 + 2{j_2}{i_3}{L_2}^2 + {i_3}^2{L_2}^2} \right)}} \tag{A--10}\\
K_{66}^{\rm{e}} &= \frac{{{i_1}\left( {3{i_1}^2 + 26{i_1}{i_2} + 15{i_2}^2} \right)}}{{(2{i_1} + {i_2})({i_1} + 6{i_2})}} \tag{A--11}
\end{align}


\end{document}